\documentclass[a4paper,fleqn]{cas-dc}

\usepackage[numbers]{natbib}

\def\tsc#1{\csdef{#1}{\textsc{\lowercase{#1}}\xspace}}
\tsc{WGM}
\tsc{QE}
\tsc{EP}
\tsc{PMS}
\tsc{BEC}
\tsc{DE}

\begin{document}
\let\WriteBookmarks\relax
\def\floatpagepagefraction{1}
\def\textpagefraction{.001}

\shorttitle{A Light Sail Astrobiology Precursor Mission to Enceladus and Europa}

\shortauthors{Lingam et~al.}

\title [mode = title]{A Light Sail Astrobiology Precursor Mission to Enceladus and Europa}                      


\author[1,2]{Manasvi Lingam}[orcid=0000-0002-2685-9417]

\cormark[1]

\ead{mlingam@fit.edu}

\affiliation[1]{organization={Department of Aerospace, Physics and Space Sciences, Florida Institute of Technology},
    city={Melbourne},
    postcode={32901}, 
     state={FL},
    country={USA}}

    \affiliation[2]{organization={Department of Physics and Institute for Fusion Studies, The University of Texas at Austin},
    city={Austin},
    postcode={78712}, 
     state={TX},
    country={USA}}

\author[3]{Adam Hibberd}[orcid=0000-0003-1116-576X]

\affiliation[3]{organization={Initiative for Interstellar Studies (i4is)},
    addressline={27/29 South Lambeth Road},
    city={London},
    postcode={SW8 1SZ}, 
    country={UK}}

\author[4,3]{Andreas M. Hein}[orcid=0000-0003-1763-6892]

 \affiliation[4]{organization={SnT, University of Luxembourg},
    addressline={29 Avenue J.F Kennedy},
    postcode={L-1855}, 
    country={Luxembourg}}

\cortext[cor1]{Corresponding author}

\begin{abstract}
Icy moons with subsurface oceans of liquid water rank among the most promising astrobiological targets in our Solar System. In this work, we assess the feasibility of deploying laser sail technology in precursor life-detection missions. We investigate such laser sail missions to Enceladus and Europa, as these two moons emit plumes that seem accessible to in situ sampling. Our study suggests that GigaWatt laser technology could accelerate a $100$ kg probe to a speed of $\sim{30}\, \mathrm{km\, s^{-1}}$, thereupon reaching Europa on timescales of $1$-$4$ years and Enceladus with flight times of $3$-$6$ years. Although the ideal latitudes for the laser array vary, placing the requisite infrastructure close to either the Antarctic or Arctic Circles might represent technically viable options for an Enceladus mission. Crucially, we determine that the minimum encounter velocities with these moons (about ${6}\,\mathrm{km\,s^{-1}}$) may be near-optimal for detecting biomolecular building blocks (e.g., amino acids) in the plumes by means of a mass spectrometer akin to the Surface Dust Analyzer onboard the \emph{Europa Clipper} mission. In summary, icy moons in the Solar System are potentially well-suited for exploration via the laser sail architecture approach, especially where low encounter speeds and/or multiple missions are desirable.
\end{abstract}



\maketitle
\section{Introduction} \label{secIntro}

Resolving the fundamental question of ``\emph{Are we alone?}'' has witnessed much progress in the 21st century \cite{SB07,RGS18,SMI18,Co20,MASD,ML21,PG21,BSSM,MA24}. In particular, a significant amount of attention has been devoted to the so-called icy worlds -- also known as ocean worlds -- in our Solar System (e.g., Enceladus, Titan, and Europa) that are empirically confirmed to harbor subsurface oceans of liquid water, which is one of the key requirements for life-as-we-know-it \cite{NP16,JL17,HHB19,HSHC,HHP21,CRK22,GBF22}. Hence, these worlds are widely perceived as promising abodes of extraterrestrial life.

Among these icy worlds, Saturn's small moon Enceladus (with radius of $\sim 250$ km) stands out by virtue of the wealth of data garnered by the \emph{Cassini-Huygens} mission \cite{LS19}. Enceladus not only hosts liquid water underneath its surface \cite{PKS09,WLM09,TTT16}, but also satisfies the other major criteria for habitability \cite{MDG18,CPG21,CTB22}, such as free energy sources \cite{HPS15,WGP17,RGW21,WMP23} and bioessential elements \cite{PKA18,KPH19,PSK23}.\footnote{Previous models suggested that dissolved phosphorus might be scarce in Enceladus' ocean \cite{MZ07,LL18}, but this prediction has been overturned by recent empirical and computational findings \cite{HGH22,PSK23,RES23}.} Furthermore, the discovery and modeling of submarine hydrothermal activity is promising \cite{HPS15,CTS17,WGP17}, because these environments are considered viable sites for engendering prebiotic chemistry and the origin(s) of life \cite{MBK08,RBB14,SSP15,BW17,TSS19,MJR21}. A number of theoretical \cite{AGS21,AGS22,HGC21,LL21,TPC21,GL23,WMP23} and experimental \cite{TPZ18,WAH22,RFH23} studies indicate that Enceladus' internal ocean might be habitable for various species of Earth-based microbes (e.g., hydrogenotrophic methanogens).

Jupiter's moon, Europa, harbors a deep subsurface ocean that ostensibly contains more liquid water than all of Earth's oceans combined \cite{Co20,ML21}. The habitability of Europa's subsurface ocean has been extensively investigated, and while not as much empirical data is available (in comparison to Enceladus), this moon is also presumed to satisfy most of the salient requirements in this respect \cite[e.g.,][]{CH01,CP01,CP02,HCP09,BH13,KP14,KK15,VHP16,RMH17,Man19,MASD,SKB20,HJV22,WMP23,MA24}. The upcoming \emph{Europa Clipper} mission \cite{HP20,VCS23},\footnote{\url{https://europa.nasa.gov/}} and the \emph{JUICE} mission to a comparatively lesser extent \cite{GDC13,OW21},\footnote{\url{https://www.esa.int/Science_Exploration/Space_Science/Juice}} will significantly enhance our knowledge of Europa's subsurface ocean. 

A striking feature of Enceladus is the existence of a substantial (time-varying) plume \cite{HES06,WCI06,TPH17,HEH19}, which extends up to $\sim 10^4$ km \cite{VHM23}. Likewise, there is evidence for plumes on Europa \cite{SHM16,SSM17,JKK18,ALS19,HRB20}, although these features may be transient \cite{RSR14,SRW19,PVR20} and/or harder to detect \cite{WH22,DOH23}. The access to plume samples would be valuable because they can shed light on the habitability of the subsurface ocean, as well as potentially enable the detection of molecular biosignatures (i.e., markers of life) \cite{MPA08,MAP14,CW19,NAD20,MNG21,NKS21,MND22}, although unraveling the latter has attendant caveats, subtleties, and lacunae \cite{MP20,EGC21,BRW22,SWW22,GJM23,MTB23,TSW23}.

A variety of putative life-detection missions to Enceladus have thus been proposed in the 2020s \cite{SWV21,DPT22,MBL22,NBC22,SGS23},\footnote{Similar recommendations have been advanced for Europa, such as the \emph{Europa Lander} mission concept \cite{HPM22}.} many of which have acquired greater relevance in light of the strong recommendation made by the comprehensive \emph{2023-2032 Decadal Strategy for Planetary Science and Astrobiology} for a flagship mission to Enceladus that would arrive at this moon in the early 2050s \cite[pg. 7]{NAS22}. The mission concept highlighted in the Decadal Strategy was the \emph{Enceladus Orbilander} concept \cite{MND21}, but alternatives such as the \emph{Enceladus Multiple Flyby} mission were also outlined. From a technological standpoint, the implementation of a plume fly-through mission is much easier than direct (i.e., in situ) exploration of the subsurface ocean \cite{DUP20}.

For missions that are geared toward sampling the plume(s), the encounter velocity of the spacecraft is rendered crucial from the perspective of identifying putative biosignatures. A multitude of publications have advocated for optimal encounter velocities in the range of around $4$-$6$ km s$^{-1}$ \cite{BPB09,KPH20a,KPH20b,JCH21,DKB23,SYC23,UMV23}, although a few studies based on other in situ techniques have posited either lower \cite{NMP20,MNG21,BAHC} or higher \cite{FSW23} speeds. Hence, mission concepts that seek to analyze the plume(s) of icy moons with subsurface oceans (i.e., ocean worlds) should ideally minimize the encounter velocity of the spacecraft with the moon and its plume(s), as this constraint could maximize the prospects of detecting putative biosignatures. 

Mission proposals to ocean worlds have hitherto relied on rocket propulsion in combination with flyby maneuvers. However, an inherent limitation of rocket propulsion is the associated mass budget and cost of the mission. In this context, propellant-free propulsion, notably light sails, could be used, which are often low thrust and can only accommodate a low-mass payload; fortunately, recent advances in hardware and software are swiftly reducing the mass and cost of the payload \cite{GC17,KHD18,OFA21,PYL21,CSW22,TCD22,ML23}. We will, therefore, explore the capacity of light sails powered by laser arrays (i.e., laser sails) -- which do not necessitate onboard propellant -- to effectuate a life-detection mission to Enceladus and Europa. The concept of light sails originated approximately a century ago \cite{Zand24,JDB29}, while laser sails were proposed in the mid 20th century \cite{Marx66,JR67}. However, research pertaining to laser sails has only taken off during the last decade \cite{Lub16,PL22a,PL22b}, driven in large part, either directly or indirectly, by the \emph{Breakthrough Starshot} initiative \cite{ZM16,PD17,ADI18,KP18,WDK18,HAH20,MLM20,ML21,DMT21,WGS21}.\footnote{\url{https://breakthroughinitiatives.org/initiative/3}}

While it is true that light sails have not been currently employed in actual deep space missions, we underscore the fact that the technology readiness level (TRL) of solar sails, which are closely associated with laser sails, is as high as $7$ to $9$ \cite{GM19,TGF23}. To give specific examples, the \emph{IKAROS} \cite{TMF11} and \emph{LightSail 2} \cite{SBB21,MBB23} spacecraft have demonstrated capabilities for achieving interplanetary missions; more recently, a prototype of a SunDiver technology demonstrator spacecraft is under development \cite{TGF23,KHT23}. Moreover, the technical details of designing and deploying solar and laser sails have witnessed extensive theoretical research \cite{McIn99,Vul12,FSE16,ADI18,KP18,GM19,WGS21,PL22a,PL22b}.

Selecting concrete targets (Enceladus or Europa) and delineating the specifics of precursor astrobiology missions to these worlds permits us to analyze and illuminate the pros (and cons) of employing laser sails in the exciting domain of carrying out extensive deep space exploration of the outer Solar System -- a new era envisioned by several publications \cite{ML21,PL22a,TKL20,SCG22,TD22,TGF23} -- which constitutes a key rationale for this paper. To the best of our knowledge, laser sail mission concepts to Enceladus or Europa have not been explicated so far despite the aforementioned benefits.

This paper assesses and subsequently demonstrates the viability of an astrobiology precursor mission to Enceladus and Europa based on a light sail architecture powered by a GigaWatt laser array. The viability is demonstrated by the identification of mission architectures where the sailcraft would encounter both bodies at relative velocities that would enable plume sampling. In Section \ref{SecMethods}, we describe the numerical code and setup used for computing optimal laser sail trajectories to Europa and Enceladus. We follow this up by presenting the major results for Europa and Enceladus in Section \ref{SecRes}. The implications of our analysis with regard to sending life-detection missions to ocean worlds by means of laser sails are outlined in Section \ref{SecDisc}. Finally, we take stock and summarize our conclusions in Section \ref{SecConc}.

\section{Methods}\label{SecMethods}

The orbital dynamics analysis herein utilizes a modified version of the MATLAB code \textit{Optimum Interplanetary Trajectory Software} (OITS) (see \cite{AH1}), and further harnesses the Non-Linear Problem (NLP) solving software MIDACO \cite{SEB09,SG10,SEG13}. The theory and structure underpinning OITS, in addition to the MATLAB code itself, can all be accessed from \cite{AH1}. Note that OITS has been successfully employed for trajectory design and analysis in a variety of Solar System contexts, ranging from interstellar objects on hyperbolic orbits to terrestrial planets like Venus \cite[e.g.,][]{HPE19,HHE20,HLE20,HH21,HPH21,HEL22,HHE22,HLH22,AH23,HA23}.

Yet, to ensure that our treatment is as self-contained as possible, we will first describe the salient assumptions and formulas underpinning OITS. First and foremost, OITS supposes that the velocity changes ($\Delta V$) exerted by thrust are impulsive, i.e., in other words these changes are instantaneous (and therefore equivalent to infinite thrust). Although this supposition is evidently applicable to high thrust propulsion schemes such as chemical rockets, nevertheless it may also be invoked for modeling the velocity change at Earth induced by a laser beam on a light sail \cite{KP18,WGS21,KP22,MLM22}.

\begin{figure}
\centering
\includegraphics[scale=0.3]{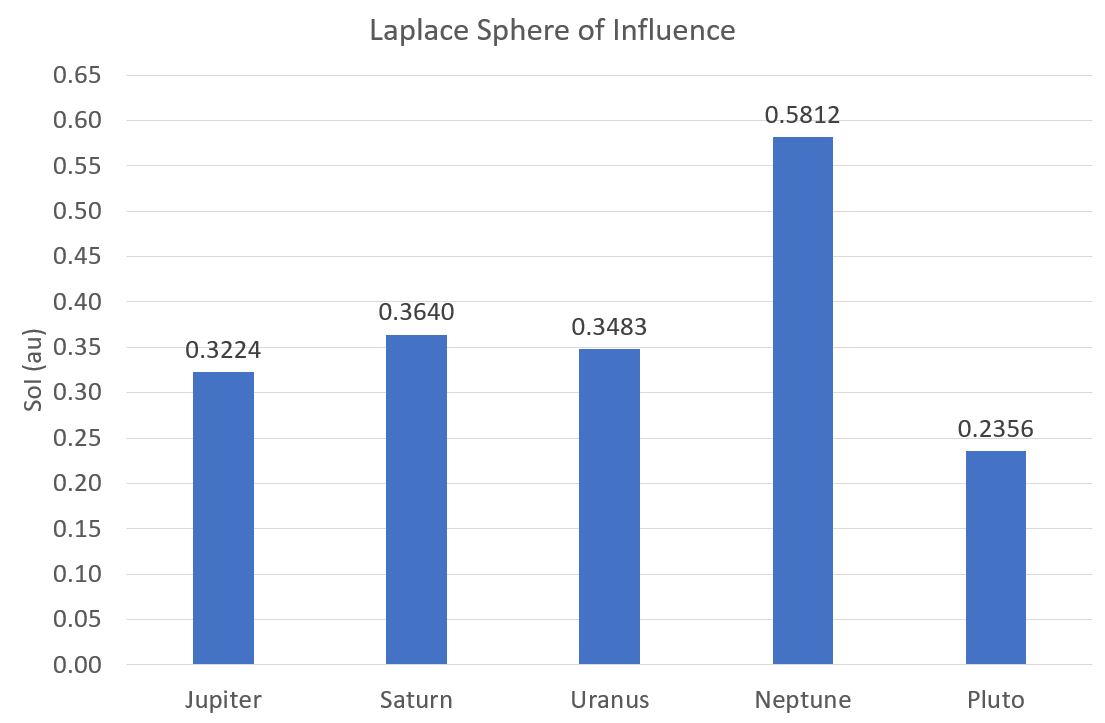}
\caption{The Laplace sphere of influence computed for the four planets of the outer Solar System, as well as for Pluto.}
\label{fig:LSOI}
\end{figure}

Second, the trajectory of the spacecraft (namely, the laser sail in this instance) is taken to be regulated exclusively by the gravitational attraction exerted by the Sun (whose gravitational parameter is $\mu_{\odot} = GM_{\odot} \approx 1.33 \times 10^{20}\, \mathrm{m^3s^{-2}}$) until the laser sail enters within the Laplace sphere of influence (SoI) \cite{DAV01,GRH22} of a user-specified planet, such as Saturn or Jupiter. In Figure \ref{fig:LSOI}, we have depicted the SoIs of the outer planets from Jupiter to Neptune, as well as the dwarf planet Pluto for the sake of contrast.

$E$ denotes the home planet (Earth) and $P$ represents the host planet orbited by the target moon under consideration. If we depart from the former at time $t_e$, and arrive at the latter at time $t_p$ (clearly, causality demands $t_p > t_e$), we can derive from these times (and given objects), an initial position vector $\vec{r}_e$ and final target position vector $\vec{r}_p$, respectively \cite{JD88}. These positions are solely expressible as a function of time, because they are calculated accurately by using appropriate interpolation functions in the NASA SPICE software library available from NAIF \cite{CHA96,ABS18},\footnote{\url{https://naif.jpl.nasa.gov/naif/toolkit.html}} and exploiting the relevant SPICE kernel files.

Now, equipped with the above $t_e$, $t_p$, $\vec{r}_e$ and $\vec{r}_p$, it is found that there exist two different solution routes connecting $E$ to $P$ via two conic arcs, which we shall christen the \textit{long way} and \textit{short way} (see \cite[Chapter 5]{HC20}). It is worth mentioning that these two transfer arcs will necessarily lie in the plane made by $\vec{r}_e$ and $\vec{r}_p$, and further that the angle $\alpha$ swept at the Sun by the laser sail on the \textit{short way} is derived from
\begin{equation}
\vec{r}_p \cdot \vec{r}_e = |\vec{r}_p||\vec{r}_e| \cos(\alpha),  
\end{equation}
where $ \vec{r}_p \cdot \vec{r}_e$ is the dot product of $\vec{r}_p$, $\vec{r}_e$ and $|\dots|$ is the magnitude of the given vector. The corresponding angle for the \textit{long way} naturally translates to $2\pi - \alpha$. This problem of determining the two trajectory solutions, \textit{long way} and \textit{short way}, between two points in space and with a given time-of-flight (in this case equal to $t_p - t_e$) is equivalent to the classic Lambert problem \cite[Chapter 7]{DAV01}. A method of solving this orbital dynamics problem is provided in \cite{Bate1971} and \cite[Chapter 5]{HC20}, and its applicability to OITS is elaborated in \cite{AH2}. Note that OITS operates under the assumption $\alpha < 2\pi$, i.e., there are no complete orbital revolutions.

Clearly, to model a laser sail encounter with, say, a moon of Saturn, the gravitational field of Saturn must be taken into account. Currently, OITS only models this field if the spacecraft happens to be in transit, i.e., using Saturn for a gravitational assist, but not if the actual target (viz. the moon) itself is within Saturn's influence. However, this issue can be addressed by harnessing a \textit{patched conic} approximation and the useful notion of an \textit{intermediate point}, defined as a point in space whose radial distance from some reference body is user-specified, but whose polar angles of longitude and latitude ($\theta$ and $\phi$) are allowed to be free and therefore can be optimized by the NLP solver in question. An extensive description of the \textit{intermediate point} approach used in this work is furnished in \cite{AH2}.

\begin{figure*}
\centering
\includegraphics[scale=0.47]{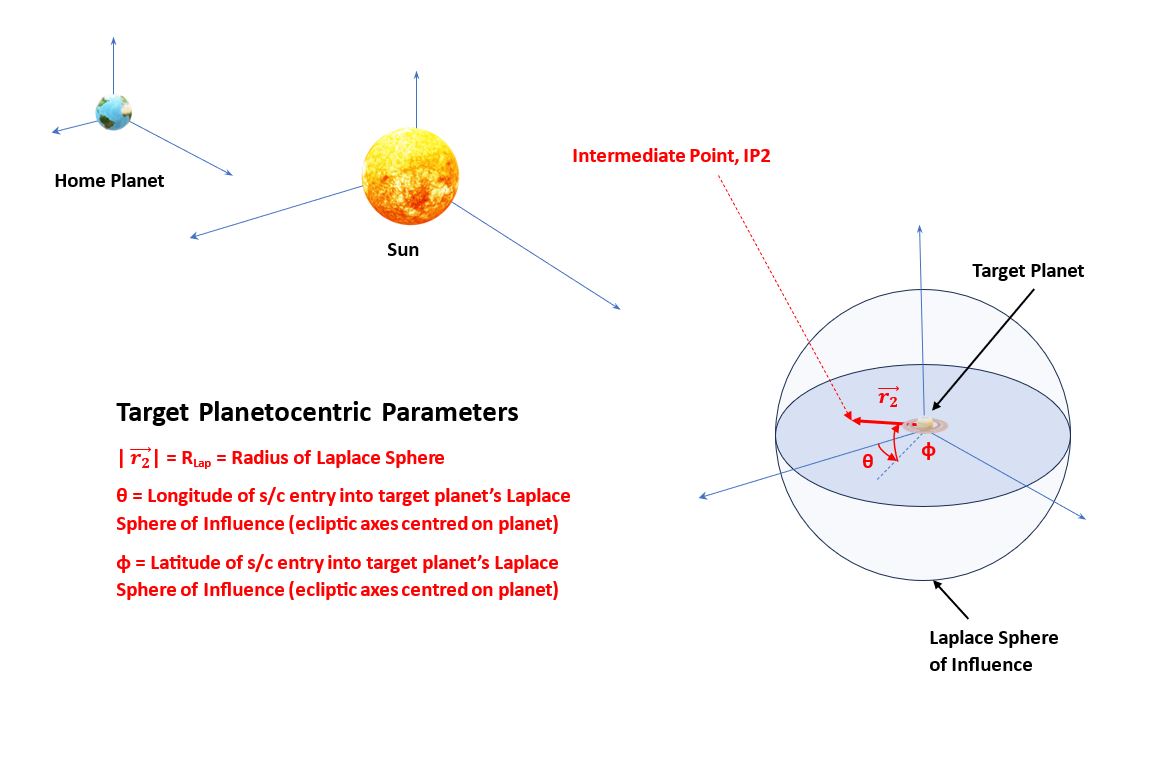}
\caption{Planetocentric definition of IP2 associated with the arrival of the light sail spacecraft (s/c) at the target planet's SoI, in conjunction with the relevant planetocentric parameters shown for the setup described in Section \ref{SecMethods}.}
\label{fig:Diag1}
\end{figure*}

\begin{figure*}
\centering
\includegraphics[scale=0.47]{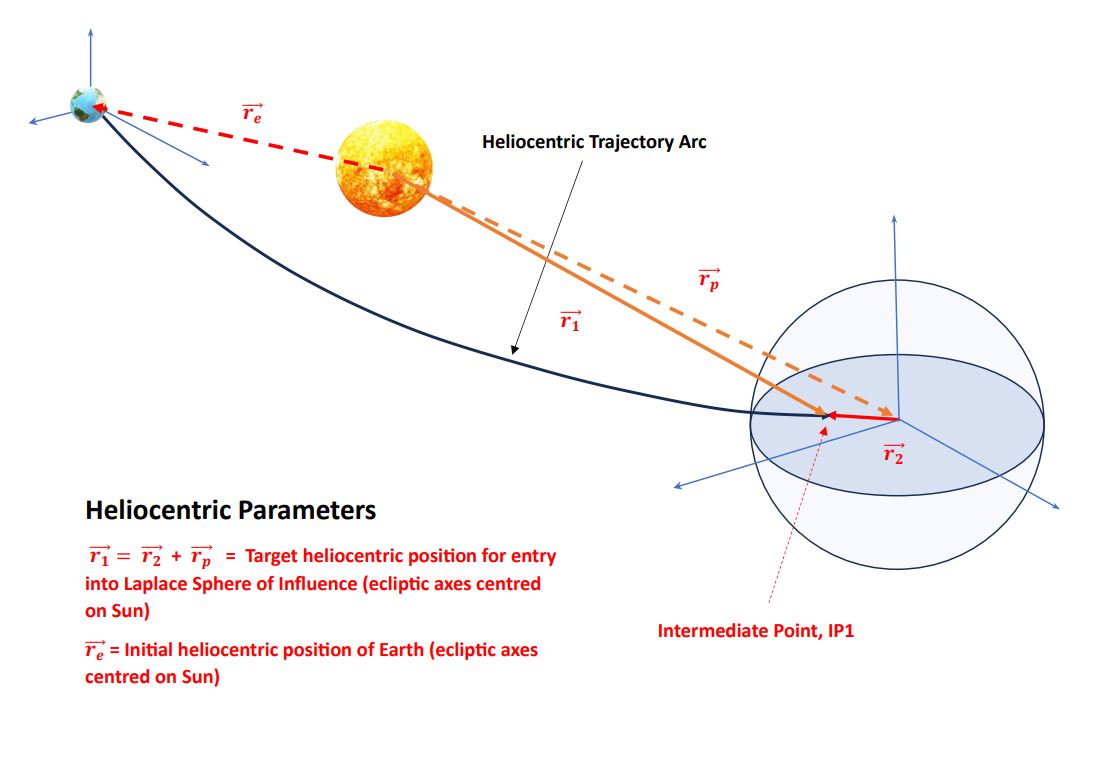}
\caption{Heliocentric trajectory associated with the light sail, in conjunction with the relevant heliocentric parameters shown for the setup described in Section \ref{SecMethods}.}
\label{fig:Diag2}
\end{figure*}

If we select the target moon's host planet (Saturn in our example) as the reference body for the \textit{intermediate point}, and take that planet's (i.e., Saturn's) SoI as its radial distance, then we end up with the scenario illustrated in Figure \ref{fig:Diag1}. We label this \textit{intermediate point} as \textit{IP2}, and we define this as the entry point of a laser sail into Saturn's SoI, having traversed the heliocentric trajectory arc from Earth, which is depicted in Figure \ref{fig:Diag2}. We remark that \textit{IP2} is defined to possess a position vector, $\vec{r}_2$, relative to the center of \textit{P}, defined by the planet's (Saturn's) SoI radius, and further by the planetocentric longitude ($\theta$) and latitude ($\phi$); the latter two serve as optimization parameters for MIDACO. 

Referring now to Figure \ref{fig:Diag2}, we witness the same trajectory from a heliocentric perspective. In this scenario, the heliocentric arrival point at Saturn's SoI is modeled via an \textit{intermediate point} labeled as \textit{IP1} (note that \textit{IP1} is the same point as \textit{IP2}, albeit in a different frame), located at heliocentric position $\vec{r}_1$ given by:
\begin{equation}
\vec{r}_1 = \vec{r}_p + \vec{r}_2, 
\label{eqcon}
\end{equation}
which follows from the simple rules of changing frames (Galilean relativity). This equation tells us that $\vec{r}_1$ is not only a function of $\theta$ and $\phi$, but also of $t_p$, with the latter specifying the position of the planet $\vec{r}_p$. The above (\ref{eqcon}) when combined with (\ref{eqcon2}) below constitute the \textit{patched conic} approximation referenced previously.

\begin{figure*}
\centering
\includegraphics[scale=0.47]{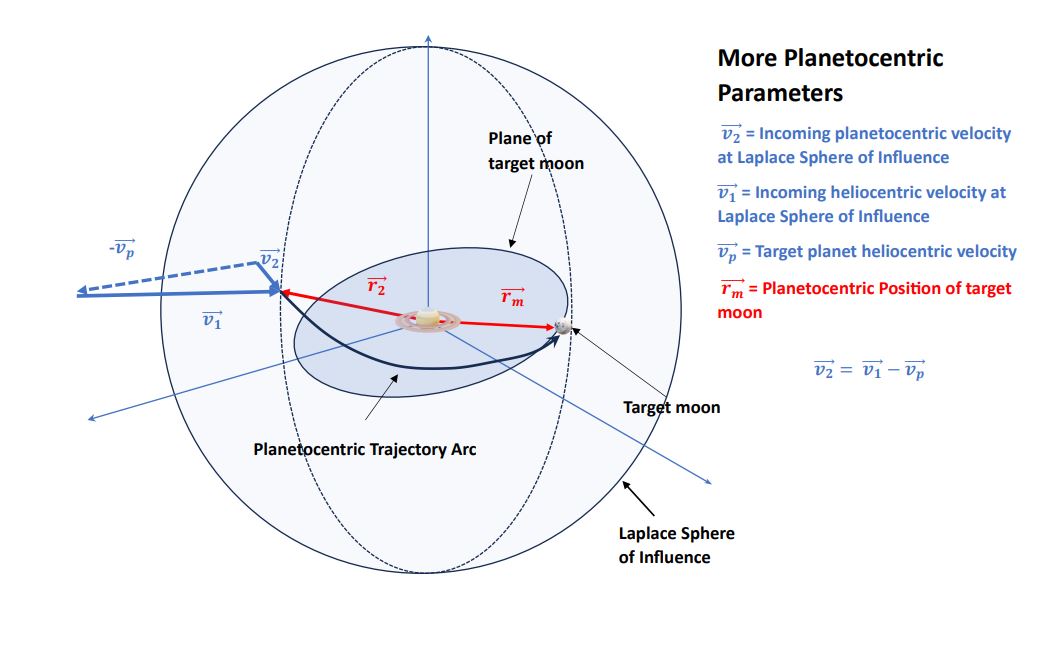}
\caption{Planetocentric trajectory associated with the light sail depicted primarily within the SoI of the planet orbited by the target moon in question, i.e., corresponding to the setup described in Section \ref{SecMethods}.}
\label{fig:Diag3}
\end{figure*}

Finally, let us refer to Figure \ref{fig:Diag3}, where we revert back to the planetocentric axis system. We must now transform the laser sail's heliocentric arrival velocity vector, $\vec{v}_1$, into the planetocentric frame, by subtracting the planet's heliocentric velocity, $\vec{v}_p$ in accordance with the rules of Galilean relativity, thereupon yielding
\begin{equation}
    \vec{v}_2 = \vec{v}_1 - \vec{v}_p,
\label{eqcon2}
\end{equation}
where $\vec{v}_2$ represents the light sail velocity relative to the planet. The subsequent planetocentric arc from \textit{IP2} at time $t_2$ to the target moon (denoted by \textit{M}) of the given planet (viz., Saturn) at time $t_M$ (where $t_M$ fully specifies the moon's position $\vec{r}_M$, along the lines stated earlier), can also be computed in a similar fashion by means of this method, which is elaborated in \cite{Bate1971}.

The optimization criterion chosen throughout this paper is to minimize the encounter velocity of the laser sail with the target moon, denoted by $V_\mathrm{ENC}$; note that the subscript `ENC' denotes 'encounter', and not 'Enceladus'. In turn, the imposition of this objective then necessarily constrains the hyperbolic excess speed to which the laser sail is accelerated when it departs Earth. The reasons for this choice of guiding objective are twofold:
\begin{itemize}
    \item The slower the laser sail moves with respect to the target moon, the more is the quantity and quality of data that might be garnered from the encounter.
    \item When $V_\mathrm{ENC}$ is lowered, the severity of fragmentation and/or destruction of any organic compounds as the laser sail travels through a plume will be reduced, and therefore these compounds may be more readily captured and characterized.
\end{itemize}
Furthermore, MIDACO permits the specification of equality constraints. Most importantly, for the problem setup tackled in this paper, this feature takes the form of a continuity constraint from the heliocentric frame to the planetocentric frame as follows. If $\vec{v}_{IP1}$ is the velocity at \textit{IP1} in the heliocentric frame and $\vec{v}_{IP2}$ is the velocity at \textit{IP2} in the planetocentric frame, then the continuity equality constraint from \ref{eqcon2} is given by
\begin{equation}
    |\vec{v}_{IP1} - \vec{v}_{IP2} - \vec{v}_p| = 0.
\end{equation}

In addition, there is one inequality constraint applied to the trajectory optimization software, namely, to constrain the periapsis point (measured relative to the host planet) of the laser sail as it approaches the moon from \textit{IP2}, to not fall below a specified lower bound. This constraint is vital because it allows the laser sail to fly past the host planet with no danger of collision, and consequently enables the probe to transmit measurements and observational data back to Earth for analysis. In this study, the minimum periapsides are conservatively chosen to be just marginally lower than the semimajor axes of the target moons in question. 

Moreover, by adopting these constraint values, rapid convergence to the optimal solution is attainable, since optimal conditions (to wit, minimum $V_\mathrm{ENC}$) are expected when the moon is at an extreme displacement from its host planet, in the plane containing the two vectors $\vec{h}_1 = \vec{r}_M \times  \vec{v}_\mathrm{s/c}$ and $\vec{h}_2 = \vec{v}_\mathrm{s/c} \times \vec{h}_1$, where $\vec{r}_M$ is the planetocentric position vector of the target moon at intercept and $\vec{v}_\mathrm{s/c}$ is the planetocentric velocity vector of the spacecraft at this moment.

It is instructive at this juncture to derive a minimum encounter velocity $V_\mathrm{ENC,min}$ of a spacecraft sent from Earth \textit{E} to the target moon \textit{M} orbiting another planet \textit{P}. To this end, we adopt a heliocentric minimum energy \textit{Hohmann transfer} \cite[Chapter 6]{HC20}, to take the spacecraft (labeled by \textit{S}) from \textit{E} to \textit{P}; on implementing the conservation of energy, we obtain
\begin{equation}
| \vec{v}_s |^2 - \frac{2\mu_{\odot}}{|\vec{r}_p|} = - \frac{\mu_{\odot}}{a},
\label{consE}
\end{equation}
where $\vec{v}_s$ is the heliocentric velocity of the spacecraft on arrival at the planet \textit{P's} SoI, and the semimajor axis $a$ in the above formula is given by $2a = |\vec{r}_p| + |\vec{r}_e|$. Upon inserting this expression into (\ref{consE}), this leads to
\begin{equation}
|\vec{v}_s|^2 = \frac{2\mu_{\odot}|\vec{r}_e|}{a|\vec{r}_p|}
\label{Vp}
\end{equation}
Converting to a planetocentric frame velocity $\vec{v}_s'$, and assuming circular orbits for the planets, we may end up with
\begin{equation}
|\vec{v}_s'| = |\vec{v}_p| - |\vec{v}_s|.
\label{Vs}
\end{equation}
Finally, on determining the minimum encounter velocity with \textit{M} -- namely, presuming that \textit{M} is situated in a circular orbit around its host planet, and moreover, supposing that the approach trajectory is coplanar with \textit{M} -- and also postulating that $\vec{v}_s'$ represents the velocity at infinity with respect to the planet, we duly arrive at
\begin{equation}
V_\mathrm{ENC,min} = \sqrt{|\vec{v}_s'|^2 + \frac{2\mu_p}{|\vec{r}_M|}} - |\vec{v}_M|,
\label{Venc}
\end{equation}
where $\vec{v}_M$ is the moon's (planetocentric) velocity. The above equation further assumes the gravitational acceleration due to the moon \textit{M} is comparatively negligible, which is indeed a valid approximation for the two moons studied herein.

\begin{figure}
\centering
\includegraphics[scale=0.28]{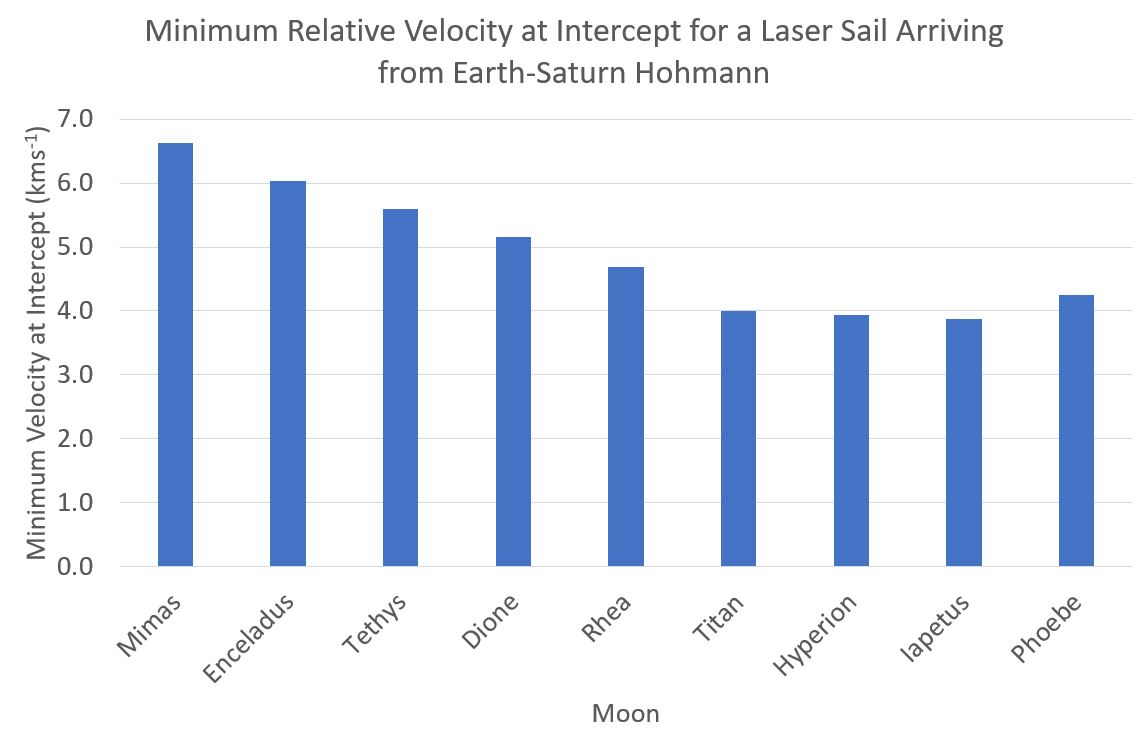}
\caption{Minimum encounter velocities feasible for laser sail missions to various major moons of Saturn.}
\label{fig:SaturnM}
\end{figure}

\begin{figure}
\centering
\includegraphics[scale=0.28]{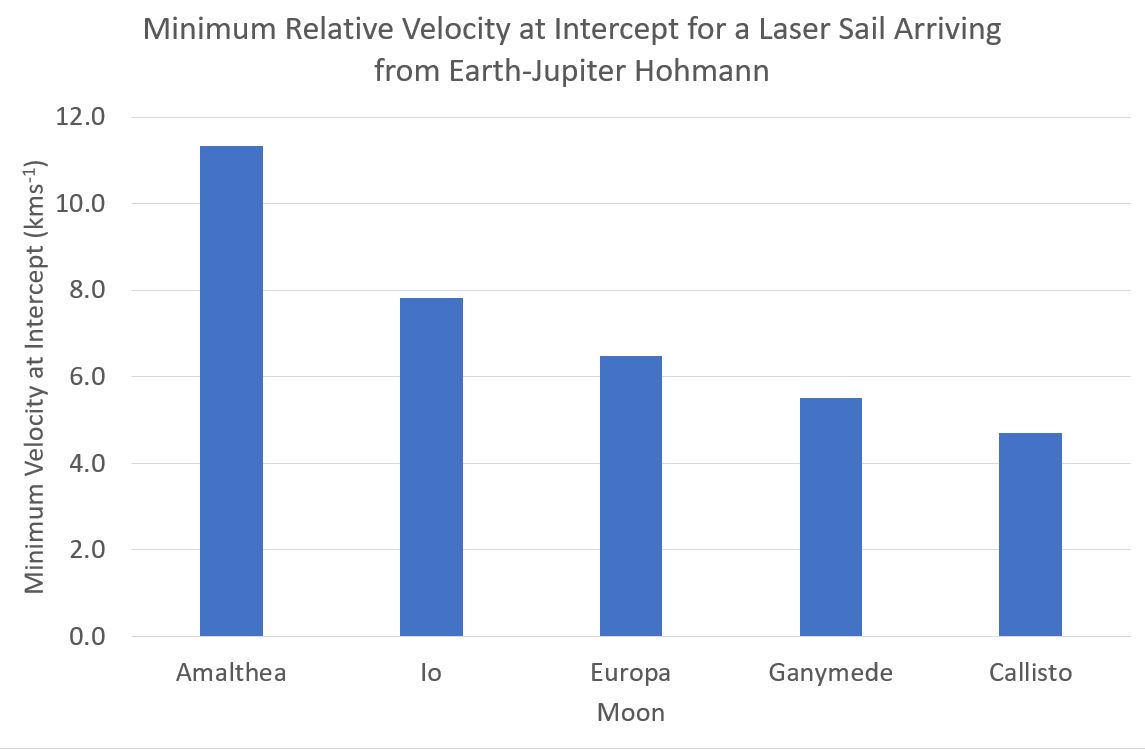}
\caption{Minimum encounter velocities feasible for laser sail missions to various major moons of Jupiter.}
\label{fig:JupiterM}
\end{figure}

The paper concentrates primarily on a potential mission to Saturn's moon Enceladus for reasons delineated in Section \ref{secIntro}. However, as a secondary consideration, we also report the results for an analogous mission to Jupiter's moon, Europa. The results ensuing from applying the preceding equations to some notable moons of Saturn and Jupiter are depicted in Figures \ref{fig:SaturnM} and \ref{fig:JupiterM}, respectively. All moons entail hypervelocity impacts, which are interpreted to be impacts that fulfill the criterion $V_\mathrm{ENC} > 3\ \mathrm{km\,s^{-1}}$. Furthermore, with $V_\mathrm{ENC} \approx {6}\ \mathrm{km\,s^{-1}}$, Enceladus and Europa are at the upper end of the velocity threshold where the sampling and detection of molecular biosignatures via in situ instruments may be viable, as elucidated in Sections \ref{secIntro} and \ref{SSecEncVel}.

\section{Results}\label{SecRes}

As indicated earlier, the analysis is conducted primarily for a putative laser sail mission to Enceladus, and secondarily for a similar mission to Europa; the pertinent orbital elements for these moons are provided in Table \ref{ORB}.

Results were generated for a specific year of launch, conservatively chosen to be 2045, and also for an entire orbital cycle of the given host planet. The latter analysis allows the full range of Earth and host planet alignments to be evaluated, and therefore an overall global minimum for this planet can be deduced. Saturn possesses an orbital period of about ${30}$ yr, based on which launch years between 2035 and 2065 were considered. For the other target, Jupiter, the shorter $\sim {12}$ yr period motivated a launch year window spanning 2040 to 2052. 

From the above paragraph, it is evident that the laser sail infrastructure is presumed to be built and operational $\sim 10$-$40$ yr in the future; for example, the fiducial date of 2045 translates to $\sim 20$ yr in the future. The required timeline of having the laser sail infrastructure assembled in a few decades is consistent with current predictions, such as the \emph{Breakthrough Starshot} initiative \cite{ZM16,PD17,WGS21}. Last, but not least, a launch date of 2045 would lead to the light sail reaching the target moon (Europa or Enceladus) circa 2050, which broadly matches the timeline prescribed for the Enceladus flagship mission recommended by the \emph{2023-2032 Decadal Strategy for Planetary Science and Astrobiology}, as outlined in Section \ref{secIntro}.

\begin{table}[]
\begin{tabular}{|c|c|c|}
\hline
\textbf{Orbital}                          & \textbf{Enceladus}    & \textbf{Europa} \\
\textbf{Parameter}                        & \textbf{(Saturn)} & \textbf{(Jupiter)}  \\ \hline
$a$ ($\mathrm{km}$)                          & 238,320            & 671,150          \\
$i$ ($^{\circ}$)                               & 28.05              & 1.88             \\
Period (days)                             & 1.37               & 3.55             \\
$e$                                         & 0.0056             & 0.0094           \\
Speed ($\mathrm{km\,s^{-1}}$)               & 12.62              & 13.74            \\ \hline
\end{tabular}
\caption{Pertinent orbital parameters for Enceladus and Europa where the reference frame is ECLIPJ2000 from SPICE that is centered on the host planet}
\label{ORB}
\end{table}

\subsection{Enceladus}

\begin{figure}
\hspace{-0.8cm}
\includegraphics[scale=0.30]{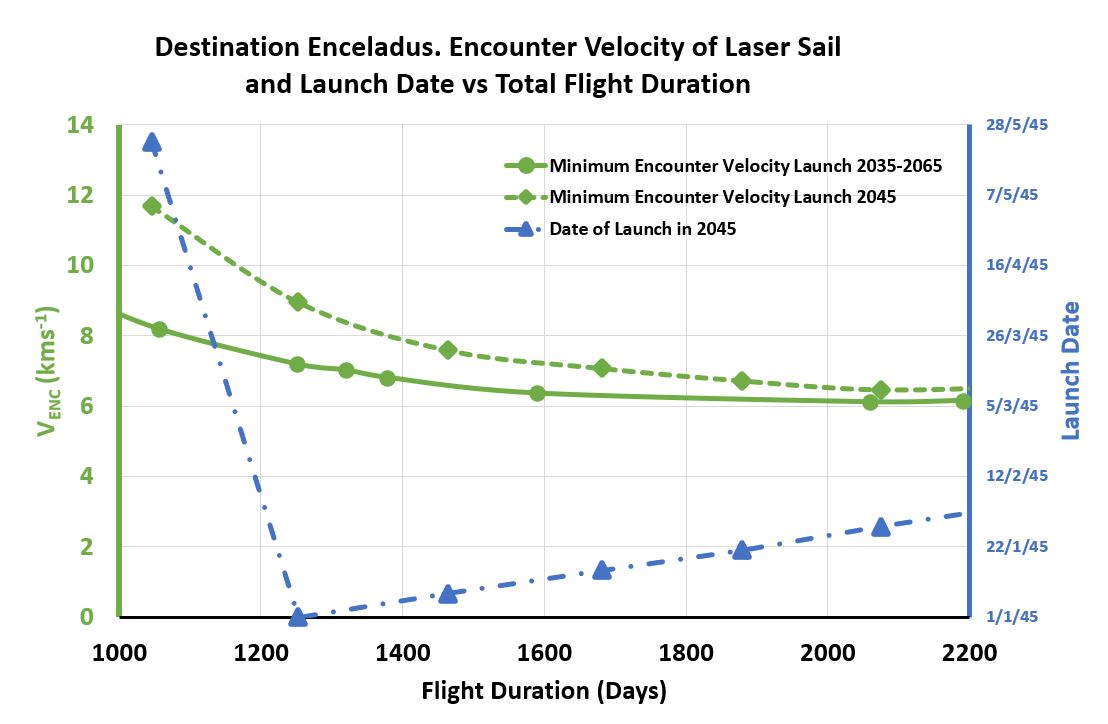}
\caption{Minimum encounter velocity (in km s$^{-1}$) of the laser sail with Saturn's moon Enceladus (left-hand axis) and the optimal launch date in 2045 (right-hand axis), versus the flight duration (in days).}
\label{fig:Enc1}
\end{figure}

The green curves in Figure \ref{fig:Enc1} illustrate the minimum encounter velocity $V_\mathrm{ENC}$ (left-hand y-axis) of the laser sail as it approaches Enceladus. The solid curve illustrates minima for launch years ranging from 2035 to 2065, whereas the dashed curve is specifically for launch dates in the year 2045; the blue dash-dotted curve indicates the corresponding launch date for the year 2045 (right-hand y-axis). In this plot, the horizontal x-axis represents the total flight duration in days. Given that a \textit{Hohmann transfer} to Saturn is known to entail a flight duration of $\sim{2200}$ days, from this plot, we ascertain that the minimum $V_\mathrm{ENC}$ would be approximately $6\,\mathrm{km\,s^{-1}}$ for this duration. This number displays excellent agreement with the rationale laid out in (\ref{consE}) to (\ref{Venc}), and the minimum $V_\mathrm{ENC}$ calculated for Enceladus in Figure \ref{fig:SaturnM}. This agreement is therefore a confirmation of the fidelity of the results generated by this software in comparison to theoretical expectations.

\begin{figure}
\hspace{-0.6cm}
\includegraphics[scale=0.30]{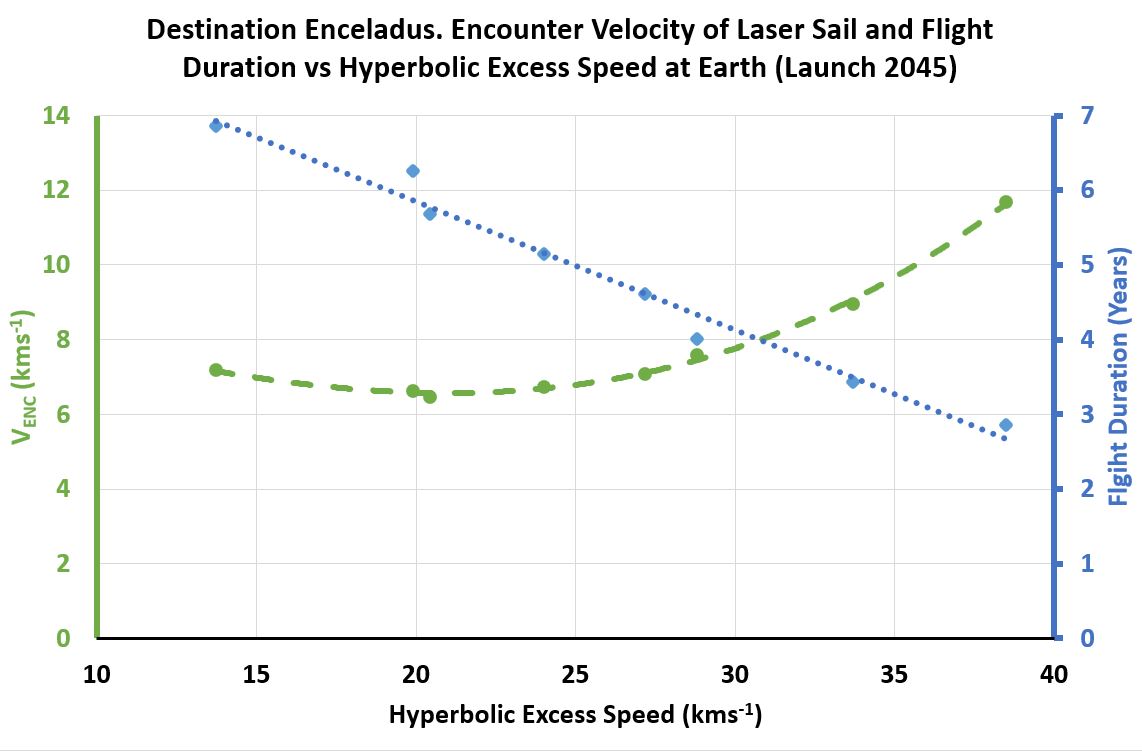}
\caption{Encounter velocity in km s$^{-1}$ (green dots and curve) and flight duration in yr (blue dots and curve) as a function of the hyperbolic excess velocity for 2045 launch date of a laser sail to Saturn's moon Enceladus.}
\label{fig:Enc2}
\end{figure}

Figure \ref{fig:Enc2} explicates the modeled mission parameters for the launch year 2045 with the x-axis now representing the hyperbolic excess speed required at Earth to reach Enceladus over a range of $V_\mathrm{ENC}$ (left-hand y-axis) and flight durations (right-hand y-axis). Note that the lowest possible $V_\mathrm{ENC}$ for this year is marginally higher than what could be achieved in theory, and is $6.4\ \mathrm{km\,s^{-1}}$ for a hyperbolic excess of $V_{\infty} = 20.4 \ \mathrm{km\,s^{-1}}$. This latter value can be used to calculate the \textit{characteristic energy} $C_3 = V_{\infty}^2$ required from the laser beam, yielding $C_3 \approx 416\,\mathrm{km^2s^{-2}}$. 

\begin{figure}
\hspace{-0.6cm}
\includegraphics[scale=0.30]{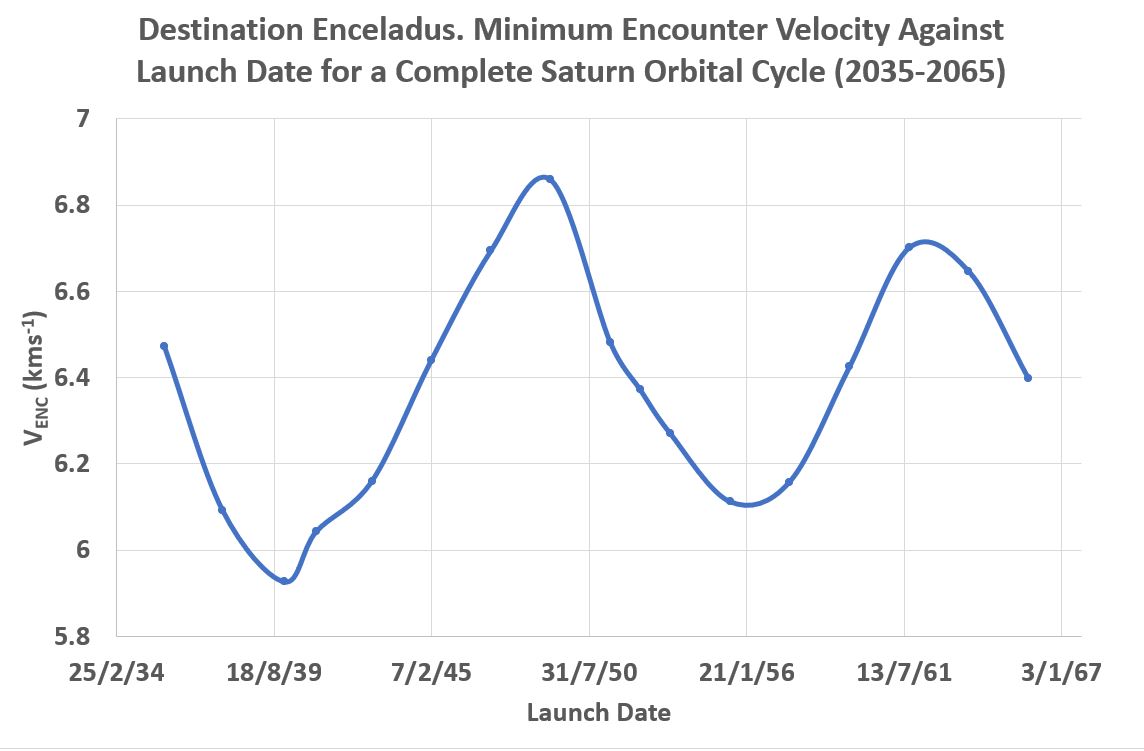}
\caption{Minimum encounter velocity in km s$^{-1}$ of the laser sail with Saturn's moon Enceladus as a function of the launch date (in the $30$ yr period 2035-2065).}
\label{fig:Enc3}
\end{figure}

We now explore how various relevant parameters evolve from year to year across the entire range of launch dates from 2035 to 2065. If, for example, we inspect Figure \ref{fig:Enc3}, the evolution of the minimum $V_\mathrm{ENC}$ over the course of this $30$ yr period is made manifest. For a launch in December 2039, $V_\mathrm{ENC}$ acquires a minimum value of $5.93\,\mathrm{km\,s^{-1}}$, which appears to be marginally lower than the minimum shown for Enceladus in Figure \ref{fig:SaturnM}. Upon further analysis of this data point, it is found that the speed of Enceladus relative to Saturn ($|\vec{v}_M|$ in (\ref{Venc})) is higher than the value used to derive Figure \ref{fig:SaturnM}, due to fluctuations in Enceladus's orbital parameters $a$ and $e$; these fluctuations are, in turn, caused by perturbing gravitational forces. The inclusion of these additional effects entirely explains the apparent inconsistency.

On consulting Figure \ref{fig:Enc3}, we witness a periodicity in the minima of $\sim 17$ yr and in the maxima of $\sim 13$ yr. These minima are found to occur in 2039 and 2056, whereas the maxima arise in 2049 and 2062. Furthermore, it should be recognized that the fluctuations in $a$ and $e$, highlighted in the prior paragraph, have a bearing on the optimal conditions for intercept of the laser sail with Enceladus, as well as how close the host planet (Saturn) is to its ascending or descending nodes with respect to the ecliptic plane, which represent approximate minimum energy (and therefore, minimum arrival velocity) conditions for a spacecraft launched from Earth.

\begin{figure}
\hspace{-0.6cm}
\includegraphics[scale=0.30]{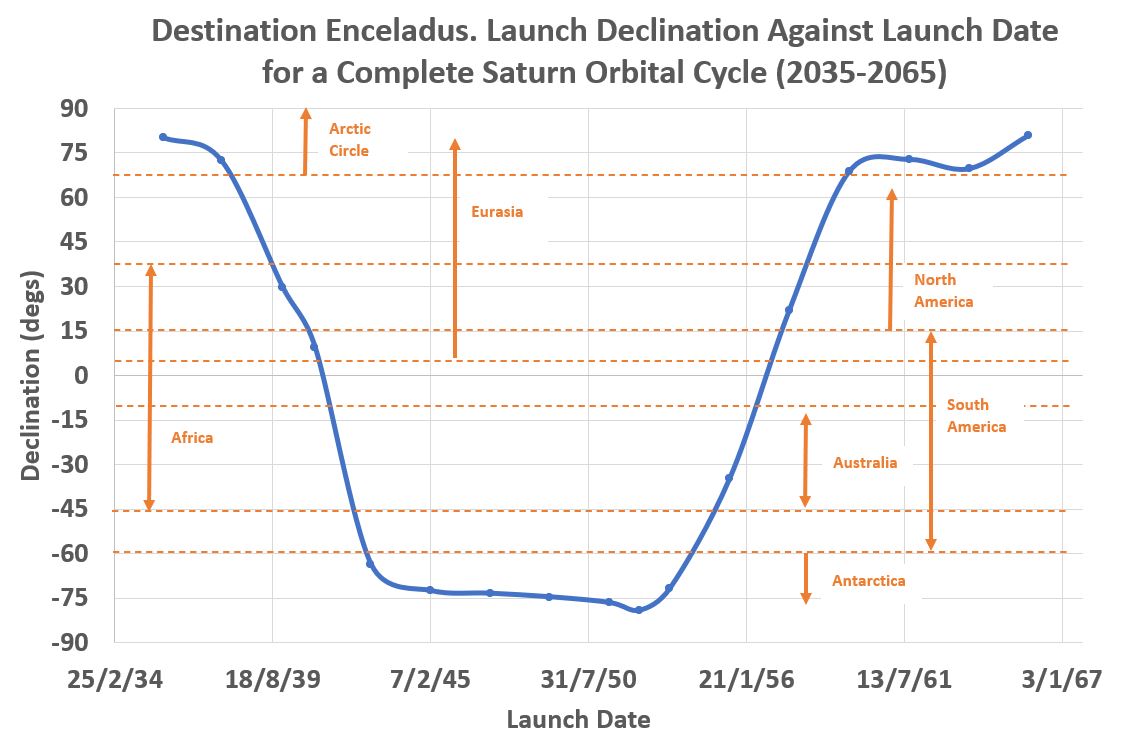}
\caption{Launch declination (in deg) and associated geographical region(s) as a function of the launch date over a complete orbital cycle of Saturn, namely, amounting to a $30$ yr period spanning 2035 to 2065.}
\label{fig:Enc4}
\end{figure}

Figure \ref{fig:Enc4} displays the dependency of the declination on the launch year, with the latter ranging from 2035 to 2065 (i.e., one complete orbital period of Saturn). The declination referenced in this plot is actually the declination of the hyperbolic excess velocity at Earth ($\vec{v}_{\infty}$), as measured in the widely utilized NASA SPICE J2000 reference frame \cite{CHA97}. The geographical region(s) on planet Earth corresponding to the computed declination are also delineated in Figure \ref{fig:Enc4}.

\begin{figure*}
\centering
\includegraphics[scale=0.31]{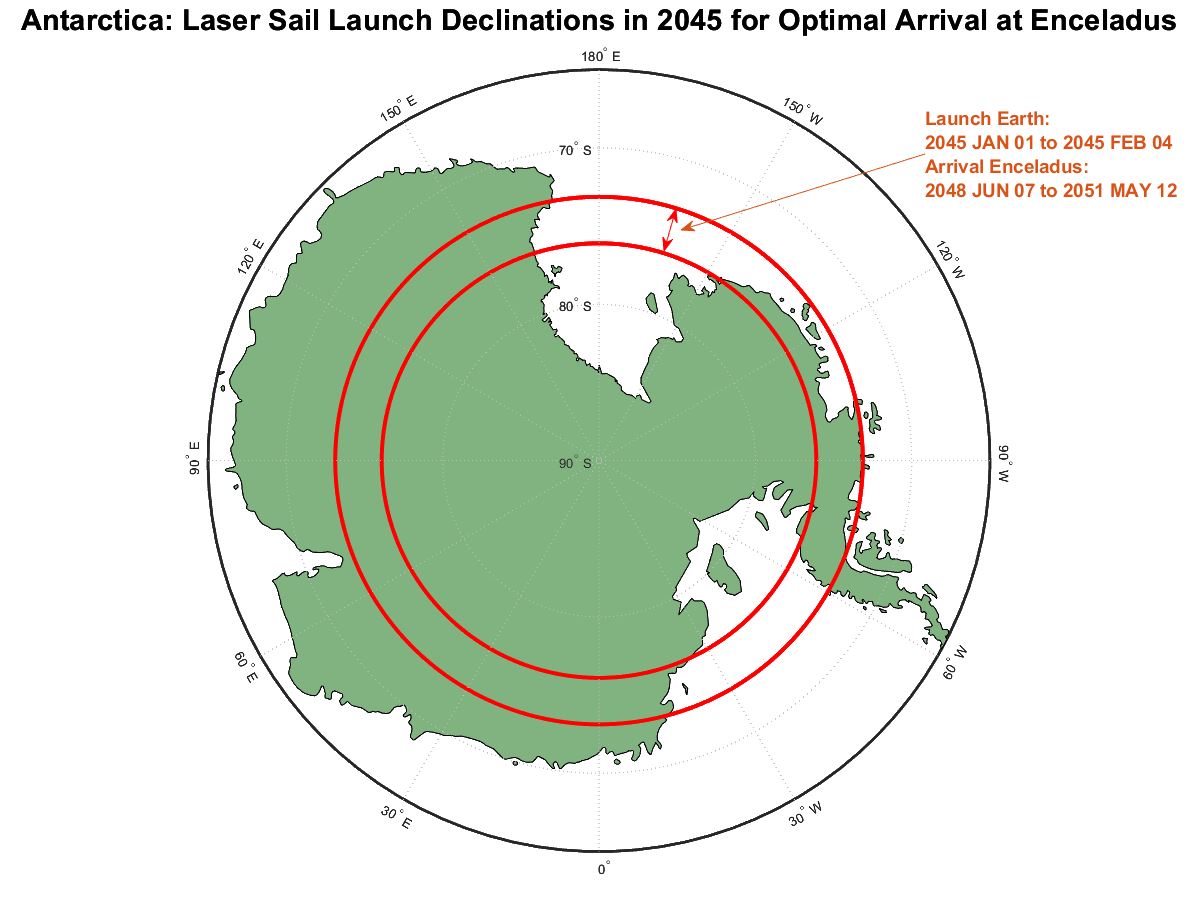}
\caption{Regions in the Antarctic compatible with achieving an optimal laser sail encounter velocity with Saturn's moon Enceladus for a 2045 launch date.}
\label{fig:Enc5}
\end{figure*}

The optimal laser array locations on Earth, in terms of providing the longest launch windows lasting approximately $9$-$10$ yr, are within the Arctic and Antarctic circles, although the minimum $V_\mathrm{ENC}$ trajectories, in 2039 and 2056 unfortunately occur at crossover points where the declination changes each year quite considerably. Although the Arctic or Antarctic seem, \emph{prima facie}, like inconvenient locations for laser sail launches, the technical advantages of locating the laser infrastructure in the Arctic or Antarctic are elucidated in \cite{PL20}.\footnote{We caution, however, that this paper does not explore the ethical or legal challenges of constructing the infrastructure in the Arctic or Antarctic, both of which are significant. This statement is also applicable to other regions foregrounded by this work.} Figure \ref{fig:Enc5} demarcates the possible locations of the laser infrastructure in the year 2045 for achieving an optimal encounter velocity with Enceladus -- in Antarctica, somewhere within the two red geographic parallels.

\subsection{Europa}
Next, we shall tackle one of the moons of Jupiter, Europa, for reasons sketched in Section \ref{secIntro}. Jupiter is endowed with an orbital period of $11.9$ yr, owing to which an entire orbital period between 2040 to 2052 was investigated for launches, as stated previously. In tandem with this long term study, a specific launch year of 2045 was chosen, to wit, the same as that adopted for Enceladus.

\begin{figure}
\hspace{-1cm}
\includegraphics[scale=0.30]{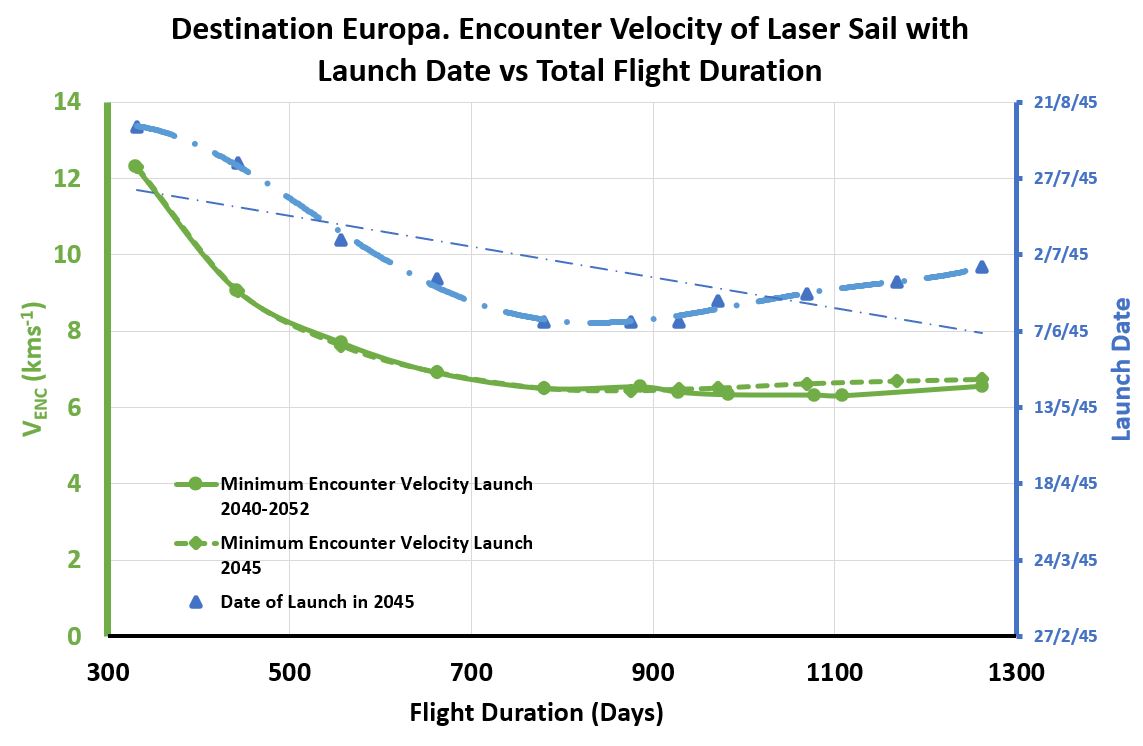}
\caption{Minimum encounter velocity (in km s$^{-1}$) of the laser sail with Jupiter's moon Europa (left-hand axis) and the optimal launch dates in 2045 (right-hand axis), versus the flight duration (in days). Note that the thin blue dash-dotted curve signifies the long-term (secular) variation of the launch date with flight duration.}
\label{fig:Eur1}
\end{figure}

\begin{figure}
\hspace{-1cm}
\includegraphics[scale=0.30]{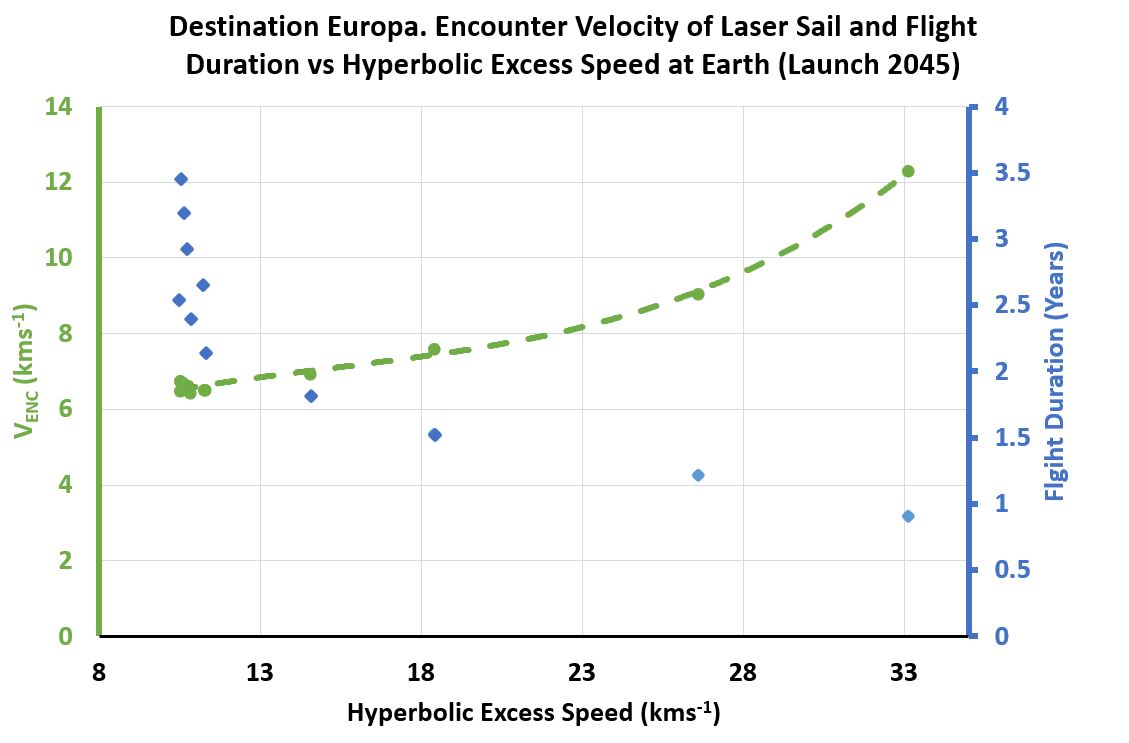}
\caption{Encounter velocity in km s$^{-1}$ (green dots and curve) and flight duration in yr (blue dots) as a function of the hyperbolic excess velocity for 2045 launch date of a laser sail to Jupiter's moon Europa.}
\label{fig:Eur2}
\end{figure}

\begin{figure}
\hspace{-1cm}
\includegraphics[scale=0.30]{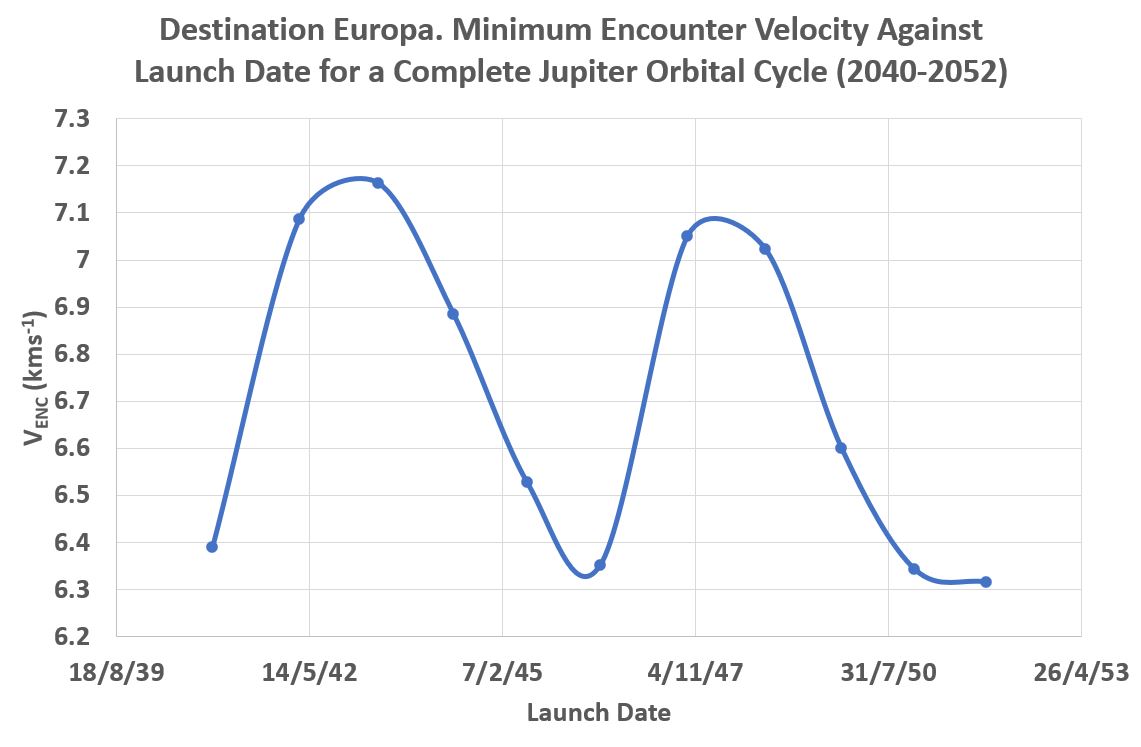}
\caption{Minimum encounter velocity in km s$^{-1}$ of the laser sail with Jupiter's moon Europa as a function of the launch date (in the $12$ yr period encompassing 2040-2052).}
\label{fig:Eur3}
\end{figure}

The salient findings are depicted in Figures \ref{fig:Eur1} and \ref{fig:Eur2}. Collectively, these plots demonstrate that 2045 is largely, albeit involving a wide range of flight durations, an optimal year for launch to Europa. This inferred result is corroborated by Figure \ref{fig:Eur3}, from which we also notice that the variations in $V_\mathrm{ENC}$ follow a $\sim 6$ yr cycle with periodicity evinced by the minima and maxima. The global minimum of about $6.3$ km s$^{-1}$ in the chosen time period displays excellent agreement with the theoretical prediction in Figure \ref{fig:JupiterM}, thereby serving to bolster the robustness of the results derived from our simulations and code.

\begin{figure}
\hspace{-1.1cm}
\includegraphics[scale=0.30]{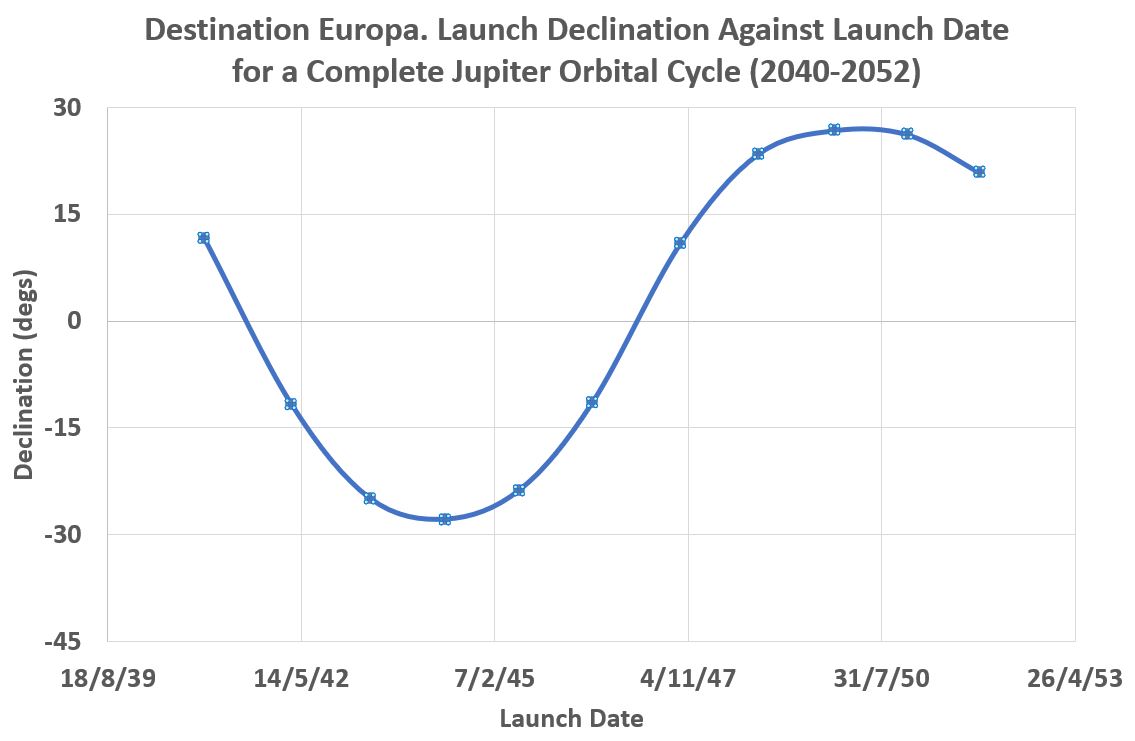}
\caption{Launch declination (in deg) as a function of the launch date over a complete orbital cycle of Jupiter, amounting to a $12$ yr period spanning 2040 to 2052.}
\label{fig:Eur4}
\end{figure}
 
The declination (of Earth hyperbolic excess velocity) against launch year from 2040 to 2052 is shown in Figure \ref{fig:Eur4}. We find a pattern akin to that of Figure \ref{fig:Enc4} for Enceladus, in the sense that a cyclical relationship is manifested. Optimal encounter velocities with Europa would occur around the peaks and troughs of declination, namely, at approximately $-30\,^\circ$ and $+30\,^\circ$, respectively. The latter includes several major countries for the putative laser architecture location: India, Pakistan, Nepal, China, Northern Sahara may all be suitable in principle.\\

Finally, a general observation with regard to both Enceladus and Europa is warranted. On account of the constraint of minimum encounter velocities, the optimal latitude for the laser infrastructure is subject to considerable variation from year to year. However, once the presumed laser sail infrastructure is installed at some location, the figures and exposition outlined hitherto suggest that an encounter velocity fairly close to the global minimum could be achieved at this location after a suitable wait period.

\section{Implications and Discussion}\label{SecDisc}

Now, we shall explicate some of the ramifications arising from the results in the preceding section.

\subsection{Encounter velocity and biosignatures}\label{SSecEncVel}

From Figure \ref{fig:Enc3}, we ascertain that the global minimum encounter velocity of a laser sail with Enceladus in the 2035-2065 interval is $V_\mathrm{E,S} \approx 5.9$ km s$^{-1}$. Likewise, from Figure \ref{fig:Eur3}, the corresponding global minimum for the encounter with Europa is $V_\mathrm{E,J} \approx 6.3$ km s$^{-1}$. In both instances, therefore, it is likely that the minimum encounter velocity achievable is around $V_0 \equiv 6$ km s$^{-1}$, which is supported by the theoretical calculations reported in Figures \ref{fig:SaturnM} and \ref{fig:JupiterM}.

It is now instructive to compare our findings with prior experimental work on detecting biosignatures in the plume(s) at different encounter velocities. Research indicates that encounter velocities of approximately $4$-$6$ km s$^{-1}$ are potentially optimal (i.e., attaining maximal sensitivity limits) for detecting biosignatures through one powerful method described below \cite{KPH20a,KPH20b,JCH21,DKB23,SYC23,UMV23}, and the above value of $V_0$ is close to the upper limit of this range. However, a few publications have suggested other techniques, and concluded that lower encounter velocities of $< 3$ km s$^{-1}$ \cite{NMP20,MNG21} or higher encounter velocities of $> 6$ km s$^{-1}$ \cite{FSW23} would be suitable.

The majority of these publications employed an empirical approach called Laser Induced Liquid Beam Ion Desorption (LILBID) \cite{KAB96,CLF,KUW22}, which mimics hypervelocity impacts of ice grains in space (akin to those in the plume) with onboard spacecraft instrumentation that consequently trigger the formation of ions. The ensuing mass spectra of these ions can be detected and analyzed by impact ionization mass spectrometers, such as the SUrface Dust Analyzer (SUDA) of the upcoming \emph{Europa Clipper} mission \cite{GKS23}.\footnote{https://europa.nasa.gov/spacecraft/instruments/suda/} Hence, the identification of molecular biosignatures is feasible, at least in principle, by current instrumentation if certain concentration and encounter velocity limits are fulfilled.

We will now delve deeper into some of the aforementioned papers to highlight the relevant findings.
\begin{itemize}
    \item LILBID experiments by \cite{KPH20a} concluded that select amino acids could be detected at molar concentrations of $\sim 1$ nM, peptides at $\sim 0.1$ $\mu$M, and fatty acids at $\sim 10$ nM; however, these results are sensitive to salinity, among other factors. The optimal encounter speeds, in the sense of maximizing sensitivity, were estimated to be $4$-$10$ km s$^{-1}$ for amino acids, $4$-$10$ km s$^{-1}$ for peptides, and $3$-$6$ km s$^{-1}$ for fatty acids. The latter belong to the important class of lipids, which might be fairly well preserved in plume materials \cite{BSB20,SWW22}.
    \item The salinity of Enceladus might be lower than that of seawater on Earth \cite{KMB22}. For the case of intermediate salinity with $0.1$ M sodium chloride, \cite{KPH20b} concluded, on the basis of LILBID experiments, that some amino acids and fatty acids might be detectable at molar concentrations of $\sim 1$ $\mu$M and $\sim 100$ $\mu$M, respectively. If the relative abundances of amino acids and fatty acids can be determined, these ratios may enable differentiating between abiotic and biotic sources for these molecules (as elucidated shortly hereafter). In addition, \cite{KPH20b} estimated that the optimal encounter velocities for amino acids and fatty acids incorporated in grains from salty environments are $5$-$10$ km s$^{-1}$ and $3$-$6$ km s$^{-1}$, respectively.
    \item Reactive molecular dynamics simulations by \cite{JCH21} and \cite{SYC23} concluded that optimal spacecraft encounter velocities should be for $4$-$6$ km s$^{-1}$ for fatty acids and amino acids enclosed in ice grains, but even at higher velocities, a small fraction of these organics were predicted to escape fragmentation. For example, at an encounter velocity of $7.5$ km s$^{-1}$, about $50\%$ of alanine, arginine, and histidine (all biological amino acids) remain intact \cite[Figure 7]{SYC23}; this result was relatively less sensitive to salt concentration. 
    \item \cite{DKB23} studied the mass spectra of biomolecules from \emph{Escherichia coli} and \emph{Sphingopyxis alaskensis} using the LILBID approach. In matrices with $0.01$ M sodium chloride, DNA nucleobases were detectable at concentrations as low as $\sim 1$ ppmw (i.e., around $10$ $\mu$M). The optimal encounter velocities were estimated to be $4$-$8$ km s$^{-1}$ for DNA, $4$-$7$ km s$^{-1}$ for metabolic intermediates (e.g., amino acids), and $3$-$6$ km s$^{-1}$ for fatty acids. However, even at higher encounter speeds of $7$-$10$ km s$^{-1}$, distinctive mass spectra of fragments derived from these biomolecules were discernible.
    \item Via the LILBID approach, \cite{NKK23} and \cite{NKH23} performed a comprehensive analysis of the mass spectra of various organic molecules as a function of the salt content associated with the ice grains. It was found that the detectability of these organics declined at higher salt concentrations, which was consistent with prior findings. The optimal encounter velocities were estimated to be $4$-$8$ km s$^{-1}$ for these organic molecules.
\end{itemize}
Based on the above exposition, it is plausible that the laser sail mission concept could achieve near-optimal encounter velocities (i.e., close to $V_0$), as viewed from the perspective of impact ionization mass spectrometry, with the plume(s) of Enceladus and Europa.

In turn, these encounter velocities might allow an impact ionization mass spectrometer (akin to SUDA) to not only identify organic molecules -- provided that their concentrations exceed appropriate thresholds for detection -- but also measure their relative abundances, thereby serving as a potential avenue for discovering molecular biosignatures; the chief reason is because biological pathways tend to preferentially synthesize or utilize some biomolecular building blocks over others, whereas abiotic processes ostensibly do not display the same degree of selectivity \cite{McK04,DNA11,DM14,GD14,CDG18,TOV20,SRM22}. Moreover, impact ionization mass spectrometry can theoretically record $10^4$ to $10^5$ ice grains in a single flyby, and discern biosignatures even if only a single grain harbors life; such sensitivity surpasses the capability of most analytical techniques \cite{KBN22}.

As a brief aside, generalizing beyond impact ionization mass spectra, it is worth underscoring that mass spectrometry, either by itself or in tandem with other analytical chemistry techniques (e.g., gas chromatography), constitutes a powerful and flexible tool for discovering myriad biosignatures in astrobiological settings \cite{AK22,SCS22,SMW22}, including those markers that may be partly agnostic to the specific biochemistry of putative extraterrestrial life \cite{CMT21,CHP23}.

\subsection{Payload, energy, and cost considerations}\label{SSecLSCons}

It is evident from Section \ref{SSecEncVel} that the identification of molecular biosignatures may be theoretically viable through impact ionization mass spectrometry. The next step involves selecting an appropriate scientific payload (going beyond a single mass spectrometer), whose mass is denoted by $M_{pl}$. For a given $M_{pl}$ and hyperbolic excess velocity of the laser sail, it is possible to obtain heuristic estimates of the total mass of the spacecraft ($M_{ls}$), the required total energy and power, and the total mission cost ($C_{ls}$).

\begin{table*}[]
\begin{tabular}{|c|c|c|c|c|}
\hline
\textbf{Mission}  & \textbf{World} & \textbf{Mission type} & \textbf{Payload mass} & \textbf{Predicted total cost} \\ \hline
Cassini-Huygens & Enceladus & flyby & $336$ kg & {$\$3.9$ billion} \\ \hline
Europa Clipper & Europa & flyby & $352$ kg & {$> \$4$ billion} \\ \hline
Enceladus Orbilander & Enceladus & orbiter-lander & $130$ kg & {$\$2.5$ billion} \\ \hline
Tiger & Enceladus & flyby & $80$ kg & {$\$636$ million} \\ \hline
Moonraker & Enceladus & flyby & $40$ kg & {$\$500$-$1000$ million} \\ \hline
ETNA & Enceladus & orbiter-lander & $197$ kg & {N/A} \\ \hline
AXE & Enceladus & flyby & $16$ kg & {$\$810$ million} \\ \hline
\end{tabular}
\caption{A synopsis of astrobiology (habitability and/or life-detection) missions to Enceladus and Europa, along with their mission type, anticipated payload mass, and total cost; the latter is specified at the time of either launching the mission or in a chosen fiducial year (such as 2022).}
\label{MissConc}
\end{table*}

Multiple mission concepts geared toward exploring Enceladus or Europa have been proposed, each endowed with their own set of objectives and accompanying scientific payloads. A handful of these concepts are summarized in the following list, and some of the salient details are furnished in Table \ref{MissConc}. Upon consulting these sources, a broad range of values for $M_{pl}$ are apparently expected, depending on the scope of the mission concept.
\begin{itemize}
    \item In the recently ended \emph{Cassini-Huygens} mission \cite{MSL02}, the scientific payload onboard the \emph{Cassini} orbiter, which comprised $> 10$ instruments, had a mass of $M_{pl} \approx 336$ kg;\footnote{Refer to Table 2.4.3 of: \url{https://sci.esa.int/documents/33648/36003/1567254416901-cassini-huygens.pdf}} the total mission cost at the time of launch was $\$3.9$ billion.\footnote{\url{https://solarsystem.nasa.gov/missions/cassini/mission/quick-facts/}} These specifications are quite close to that of the \emph{Europa Clipper} mission, with $M_{pl} \approx 352$ kg, a spacecraft dry mass of $\sim 2600$ kg, and a cost of $>\$4$ billion.\footnote{\url{https://europa.nasa.gov/system/internal_resources/details/original/116_Europa-Clipper-Newsletter-External-vol2issue1_FINAL.pdf}}
    \item The comprehensive orbiter-lander \emph{Enceladus Orbilander} mission concept was proposed by \cite{MND21,MND22} to explicitly address the question of whether Enceladus' subsurface ocean contains life.\footnote{A detailed breakdown of the mission concept is provided in: \url{https://ntrs.nasa.gov/citations/20205008712}} With a total of $13$ scientific instruments (a high-resolution mass spectrometer among them), the payload mass was estimated to be $M_{pl} \approx 130$ kg. The dry mass of the spacecraft was predicted to be $\sim 2800$ kg, and the mission cost (excluding that of the launch vehicle) was determined to be about \$2.5 billion in 2025.
    \item The flyby \emph{Tiger} mission concept was proposed by \cite{SWV21} to assess the habitability of Enceladus' ocean by sampling the plume to detect organics. The four instrument payload, including a mass spectrometer, amounted to $M_{pl} \approx 80$ kg. The dry mass of the spacecraft was $\sim 1000$ kg, and the predicted cost was \$636 million in 2022.
    \item The flyby \emph{Moonraker} mission concept was proposed by \cite{MBL22} to gauge the habitability of Enceladus' ocean, the mechanisms underpinning the South Polar Terrain, and the formation of the moon. A suite of eight instruments, with an ion and neutral mass spectrometer, was estimated to yield $M_{pl} \approx 40$ kg. The spacecraft had total dry mass of $\sim 1250$ kg, and a cost in the range of $\sim \$500$-$1000$ million in 2022.
    \item The orbiter-lander \emph{Enceladus Touchdown aNalyzing Astrobiology} (ETNA) mission concept was proposed by \cite{DPT22} to search for (pre)biotic signatures in the plume and assess the habitability of Enceladus' ocean. A total of seven scientific instruments, including a laser desorption/ablation mass spectrometer, had a mass of $M_{pl} \approx 197$ kg (of which the majority was a radio science instrument). The lander and orbiter had a dry mass of $\sim 100$ kg and $\sim 550$ kg, respectively. 
    \item The flyby \emph{Astrobiology eXploration at Enceladus} (AXE) mission concept was proposed by \cite{SGS23} to seek out molecular biosignatures, and perform related geophysical investigations. A minimal payload mass of $M_{pl} \approx 16$ kg was selected, consisting of a quadrupole ion trap mass spectrometer, among other instruments. The dry mass of the spacecraft could be $\sim 1400$ kg, and the predicted cost was \$810 million in 2022.
\end{itemize}

As implied earlier, the scientific payload mass spans more than one order of magnitude, which is not surprising in view of the diversity of these missions concepts. Our goal in this study is to focus on a relatively simple precursor astrobiology mission, albeit with the express understanding that this analysis can be readily generalized to more complex missions (with extended instrumentation) by means of boosting $M_{pl}$ and the other model parameters accordingly.

We adopt a fiducial low-end mass of $M_{pl} \approx 16$ kg, identical to that of the AXE mission concept \cite{SGS23}. Our rationale is that we focus on a precursor astrobiology mission, which is therefore meant to serve as a proof-of-concept. Note that this payload is sufficiently higher than the $\sim 5$ kg mass of SUDA \cite{KAB14}, whose importance was elucidated in Section \ref{SSecEncVel}. Hence, other instruments (e.g., camera; nephelometer) can be incorporated with the goal of characterizing habitability and/or garnering contextual information. Breakthroughs in portable mass spectrometers could bring down the mass further without compromising functionality \cite{YRC22}, or enhance sensitivity for the same mass \cite{FSW22}.

We will assume that the total mass of the payload (i.e., sans the sail itself) is $M_{ls} \sim 100$ kg for the above configuration, thereby warranting classification as a SmallSat \cite{YWB21}. The posited spacecraft may host, \emph{inter alia}, an $8$ kg avionics system for command and data handling, a $\sim 10$ kg Radioisotope Thermoelectric Generator for power,\footnote{\url{https://link.springer.com/article/10.1007/s11214-019-0623-9}}, and a $<10$ kg high-gain antenna designed for the outer Solar System \cite{CEC20}. We note that our choice of $M_{ls}$ is consistent with \cite[Equation 1]{GO13}, where we have used a lunar-type mission scaling in lieu of an interplanetary mission because the latter has additional mass budget for subsystems associated with (interplanetary) propulsion, which are neglected in the laser sail architecture. Moreover, if the spacecraft were to be launched in the 2040s, it is plausible that the rapid pace of miniaturization will lower the necessary mass \cite{IB03,KL18,LBB18,HBE19,CBM20,YWB21}.

As another point of comparison, although not a life-detection mission, the \emph{New Horizons} spacecraft had a $30$ kg payload and a $400$ kg dry mass \cite{AS08}; its total cost was approximately \$780 million.\footnote{\url{https://www.planetary.org/space-policy/cost-of-new-horizons}} Interestingly, the above values for the scientific payload and spacecraft mass are close to that of the \emph{Pluto Kuiper Express} mission concept from the 1990s, which involved a $9$ kg payload and a $175$ kg spacecraft \cite{AS08}. To reiterate the theme from the prior paragraph, advances in material science \cite{LBB18}, communications \cite{SEA20}, and scientific instrumentation are swiftly driving down the mass requirements.

Next, we turn our attention to the final launch speed $v_i$ of the laser sail from Earth. On inspecting Figures \ref{fig:Enc2} and \ref{fig:Eur2}, we determine that the encounter velocity at Enceladus and Europa are minimized for a hyperbolic excess velocity of $v_\infty \approx 20$ km s$^{-1}$ and $v_\infty \approx 10$ km s$^{-1}$, respectively. From the conservation of energy, we have
\begin{equation}
    v_i^2 = v_\infty^2 + \frac{2 \mu_\oplus}{R_\oplus},
\end{equation}
from which we conclude that $v_i \lesssim 30$ km s$^{-1}$ after consulting Figures \ref{fig:Enc2} and \ref{fig:Eur2}. Hence, to leading order, we approximate the velocity to which the laser sail must be accelerated by $v_i \approx 10^{-4}\,c$, where $c$ is the speed of light.

Equipped with $M_{ls} \sim 100$ kg and $v_i \approx 10^{-4}\,c$ as the inputs, we invoke the cost-optimized laser sail model of \cite{KP18}, whose key findings are depicted in \cite[Table 2]{TKL20} and \cite[Figure 7]{KP22}, with the essential condition of grid-based power supply. Some of the major results of this model, drawn from these two references, are tabulated in Table \ref{CostLS}. We are now in a position to obtain a heuristic estimate of the cost attributable to this precursor astrobiology mission. Before doing so, we emphasize that the substantial mass and low velocity of the spacecraft jointly aid in suppressing the hazards (e.g., impacts by gas and dust; electromagnetic forces; hydrodynamic drag) that would otherwise prove significant for extremely lightweight (viz., gram-mass), relativistic laser sails \cite{HLB17,HL17,LE18,HL20,LL20,LCE22,KL23}.

\begin{table}[]
\begin{tabular}{|c|c|}
\hline
\textbf{Sail parameter}  & \textbf{Estimate} \\ \hline
Sail diameter & $\sim 100$ m \\ \hline
Sail mass & $\sim 20$ kg \\ \hline
Peak radiated power & $\sim 10^9$ W \\ \hline
Capital expenditure & $\sim \$1$ million kg$^{-1}$ \\ \hline
Payload energetics cost & $\sim \$2$ million kg$^{-1}$ \\ \hline
\end{tabular}
\caption{Approximate laser sail parameters for a precursor astrobiology mission to Enceladus or Europa with a $100$ kg spacecraft (sans the sail apparatus) with launch speed of $10^{-4}\,c$. The last two rows are expressed as cost per unit mass of the spacecraft for reasons indicated in the text.}
\label{CostLS}
\end{table}

To leading order, the total cost can be crudely written down as the sum of the capital expenditure; the energy costs of launching the spacecraft (taken to be $\sim 100$ kg); and the costs of research and development, production, and testing associated with the spacecraft.\footnote{The scientific instruments onboard the light sail are conservatively identical to those from the 2010s/2020s, and are therefore presumed to entail a much lower cost in the 2040s \cite{IB03}.} The first two terms are extracted from the last two rows of Table \ref{CostLS} (which display cost per unit mass), and we nominally adopt $\sim \$0.2$ million kg$^{-1}$ for the last term of the trio \cite[cf.][Figure 2-2]{SWH16}; this value is lower than certain 2010 SmallSat technology by a factor of $2$-$3$ \cite[Table 5]{WER11}, and might thus be realizable in the 2040s. On combining these terms together for the launch of a \textit{single} light sail, we end up with
\begin{equation}\label{Cost1LS}
    C_{ls} \sim \$ 3.2\,\mathrm{million}\,\left(\frac{M_{ls}}{1\,\mathrm{kg}}\right),
\end{equation}
but if we consider a scenario wherein $\mathcal{N}_{ls} \geq 1$ light sails are launched, the above formula is modified to
\begin{equation}\label{CostNLS}
    C_{ls} \sim \$ 1\,\mathrm{million}\,\left(1 + 2.2\,\mathcal{N}_{ls}\right)\left(\frac{M_{ls}}{1\,\mathrm{kg}}\right).
\end{equation}
On substituting $M_{ls} \sim 100$ kg in (\ref{Cost1LS}), we obtain a total cost of around $\$320$ million. If we compare this result with the proposed missions delineated previously, especially the AXE mission concept (which has similar parameters), it might be feasible that the laser sail mission could be executed at a total cost that is a few times lower than conventional designs based on chemical propulsion. However, one noteworthy caveat that we shall touch on shortly is that the laser sail only permits a single flyby, whereas the concepts summarized earlier featured multiple flybys and/or landings. 

A couple of avenues whereby $C_{ls}$ in (\ref{CostNLS}) may be reduced are plausible. First, the capital and operating expenditures were derived on the basis of extrapolated rates for grid power, infrastructure, and so forth, which are at a nascent stage vis-\`a-vis light sail architectures. The drop in prices arising from maturation of such technologies would lead to a corresponding reduction in the total mission cost. We will briefly elaborate on this theme with regard to improvements in laser technology and dropping prices in this area.

The (ground-based) laser array contributes to the capital expenditure and energetics cost in Table \ref{CostLS}, constructed from \cite[Figure 7]{KP22} and \cite[Table 2]{TKL20}. An extrapolation of current laser price trends suggests that a scaling of $\lesssim \$0.1$ W$^{-1}$ is attainable by the end of the 2030s (\cite[Figure 2]{WGS21} and \cite[Figure 4-4]{ADM22}); this value may drop further to $\sim \$0.01$ W$^{-1}$ for GigaWatt technology in the 2040s \cite{WGS21,ADM22}. For a peak power of $\sim 10^8$ W -- required for a $100$ kg payload, as per Table \ref{CostLS} -- we are led toward a laser cost of $\sim \$5$--$50$ million, to wit, less than the estimated total cost of $\$320$ million, thus serving as a consistency check. Likewise, this laser cost is smaller than the capital expenditure and the energy cost for launching the $100$ kg payload, as calculated from Table \ref{CostLS}, along expected lines.

However, the above extrapolation might be conservative, because it does not fully account for the swift pace of crucial breakthroughs in laser technology \cite{JLT13,ZC14,LMS17,DCB18,ZL22}. To highlight just a representative example, photonic-crystal surface-emitting lasers (PCSELs) are single-mode semiconductor lasers capable of high power and high beam-quality \cite{SN10,NKO17,IDN19}. PCSELs can deliver a laser power up to $1$ kW from an emitter of $<10$ mm in size \cite{IYG22}, and have empirically achieved a brightness of $\sim 1$ GW cm$^{-2}$ sr$^{-1}$ \cite{NIY23}. Hence, technological advances in individual lasers (e.g., PCSELs), phase sensing and control \cite{AF19,GRS20,BSI21,HRJ23}, among several other factors, may render laser sail propulsion both realistic and affordable. This is an active area of research that goes beyond the scope of this work.

Second, by leveraging the fact that spacecraft miniaturization is proceeding apace -- in conjunction with innovative developments in autonomous vehicles, space robotics, machine learning,\footnote{In the context of machine learning, these breakthroughs can be effectively harnessed to detect agnostic molecular biosignatures \cite{CHP23}.} and onboard computing \cite{GC17,KHD18,KL18,AF20,LHJ20,OFA21,PYL21,CSW22,TCD22,TSG22,ML23,WLM24} -- it is conceivable that life-detection missions could be implemented at progressively smaller payload and spacecraft masses in the decade(s) to come, irrespective of the type of propulsion (e.g., chemical or light sail). In this case, as revealed by (\ref{Cost1LS}) and (\ref{CostNLS}), the mission cost would decline commensurately. We now illustrate this point quantitatively below.

CubeSats are already under development or evaluation for interplanetary missions \cite{SBH13,SKW17,JKR18,BBB19,CBM20}, and their viability for boosting our knowledge of astrobiology appears promising \cite{EML20,RC22,CTL23,MLS23}. Let us consider a typical 1U CubeSat with a mass of $M_{ls} \approx 1$ kg, and suppose that $\mathcal{N}_{ls} \approx 1000$ of them are launched toward an ocean world (e.g., Enceladus). On substituting these parameters in (\ref{CostNLS}), we arrive at a total cost of roughly $\$2.2$ billion, i.e., which is lower than the \emph{New Horizons} mission. Since each one of these CubeSats can perform a single flyby in principle, about $1000$ flybys ought to be theoretically possible at this cost. 

Hence, even though the science return from a single CubeSat will probably be low (i.e., limited by the onboard instrumentation), the cumulative return should be considerable because of the sheer number of spacecraft at play. A bevy of publications have therefore advocated a flotilla of small spacecraft to expand space exploration \cite{CM12,HCM16,LKC18,NT20,TBB21,TT21,SCG22,ALS23}. Last, but not least, the high number of CubeSats enables the incorporation of a diverse portfolio of instruments (e.g., different instruments on each spacecraft) and/or building significant redundancy (i.e., many spacecraft carrying similar instruments). On the whole, it is thus not an exaggeration to contend that the deployment of small spacecraft propelled by laser arrays may help revolutionize our understanding of the Solar System.

\subsection{Combining laser sails with onboard (chemical) propulsion}

As mentioned a few paragraphs above, one of the disadvantages of the laser sail mission concept is that each spacecraft would accomplish a single flyby (i.e., encounter) with the plume(s) of Enceladus or Europa at a given velocity. While this limitation does exist, it is one that can be overcome by including a suitable onboard propulsion system, since we have effectively assumed so far that no such propulsion system has been incorporated.

If an onboard propulsion system is available, then the relative velocity during the encounter can be reduced from $V_\mathrm{ENC}$ to $V_\mathrm{ENC} - \Delta v$, where $\Delta v$ is supplied by the propulsion system. In an extreme case, after setting $\Delta v \approx V_\mathrm{ENC}$, we could potentially place the spacecraft in orbit around the target moon \cite[Chapter 8]{HC20}. Note that the addition of propulsion increases the mass budget in accordance with the classic rocket equation \cite[e.g.,][]{DM19,UW19}, 
\begin{equation}\label{RockEq}
    \frac{\Delta v}{v_\mathrm{ex}} \approx \ln\left(\frac{M_{ls}'}{M_{ls}}\right),
\end{equation}
where $v_\mathrm{ex}$ is the effective exhaust velocity, $M_{ls}'$ is the new total mass (namely, the initial mass at the time of launch), while $M_{ls}$ is the original spacecraft mass introduced in Section \ref{SSecLSCons}, and constitutes the de facto final mass. Once the new mass $M_{ls}'$ has been determined, the results in Section \ref{SSecLSCons} must be updated, with $M_{ls}'$ substituted in place of $M_{ls}$, such as in both (\ref{Cost1LS}) and (\ref{CostNLS}).

As an example, let us choose N$_2$O$_4$/MMH (monomethylhydrazine) propellant, which has a fairly high boiling point (thus mitigating the amount of cryogenic equipment) and substantial exhaust velocity of $v_{ex} \approx 3.3$ km s$^{-1}$;\footnote{\url{http://www.astronautix.com/n/n2o4mmh.html}} and $\Delta v$ is specified to be $V_0$ from Section \ref{SSecEncVel} (i.e., the near-minimum encounter velocity sans onboard propulsion). After substituting these values into (\ref{RockEq}), we end up with $M_{ls}' \approx 6.16\, M_{ls}$. Therefore, for a $M_{ls} \sim 100$ kg spacecraft (this mass was motivated earlier), invoking (\ref{Cost1LS}) and then replacing $M_{ls}$ with $M_{ls}'$ translates to a total mission cost of about $\$2$ billion, which is distinctly higher than the flyby mission designs outlined in Section \ref{SSecLSCons}, but is approximately equal to the cost of either past/current flagship missions or proposed orbiters/landers.

To round off our discussion, we note that the arguments in the closing paragraphs of Section \ref{SSecLSCons} are still valid. To recapitulate, if $M_{ls}$, and consequently $M_{ls}'$ from (\ref{RockEq}), are reduced due to technological progress in miniaturization, in principle we could deploy a flotilla of laser sails to ocean worlds for analyzing their habitability, which may greatly boost the ensuing cumulative scientific return. For instance, if we adopt $M_{ls} \approx 1$ kg for a canonical 1U CubeSat as before, then we obtain $M_{ls}' \approx 6.16$ kg from the above paragraph. In turn, we calculate that $\mathcal{N}_{ls} \sim 184$ for a total mission cost of $\$2.5$ billion, after utilizing (\ref{CostNLS}) and swapping $M_{ls}$ with $M_{ls}'$. In other words, nearly $200$ CubeSat-like spacecraft might be placed in orbit around Europa and/or Enceladus for the same cost as the \emph{New Horizons} mission.

\section{Conclusion}\label{SecConc}

Ocean worlds (i.e., icy worlds with subsurface oceans) rank among the most promising targets for life-detection missions in the future because they fulfill most/all of the core criteria for habitability. Moreover, plumes on Enceladus and Europa suggest that it may be feasible to sample their interior and possibly search for molecular biosignatures. Hence, a plethora of such astrobiological missions to Enceladus and Europa have been proposed, such as the \emph{Enceladus Orbilander} with an arrival date in the early 2050s; all these missions are powered by chemical propulsion.

However, alternatives to chemical propulsion are garnering attention, notably light sails driven by solar radiation (solar sails) or lasers (laser sails). Light sails offer two major advantages: (1) lower mass and cost requirements owing to the potential absence of onboard fuel; and (2) faster flight times, because they can attain high terminal speeds. Furthermore, with advances in hardware (e.g., robotics) and software (e.g., artificial intelligence), future spacecraft are likely to become progressively miniaturized, thus permitting flotillas of low-mass spacecraft to survey worlds in our Solar System in a cost- and/or energy-efficient fashion; this modality is well-suited for light sails.

Motivated by the preceding two paragraphs, this paper focused on the icy moons Enceladus and Europa, and investigated whether an astrobiology precursor mission could survey these ocean worlds through a laser sail architecture. The following key features emerged from our analysis of optimal laser sail trajectories to Enceladus and Europa (with an emphasis on the former):
\begin{itemize}
    \item Relatively short transit intervals of $3$-$6$ years are achievable for missions to Enceladus, while the equivalent timescales for Europa are $1$-$4$ years.
    \item The minimum encounter velocity of the spacecraft with the plume(s) is predicted to be around $6$ km s$^{-1}$ for both Europa and Enceladus. Interestingly, this speed is close to the upper range of the optimal velocity necessary for detecting biomolecular building blocks (e.g., amino acids and lipids) via impact ionization mass spectrometers akin to SUDA onboard \emph{Europa Clipper} mission.
    \item Several locations are technically suitable for the light sail infrastructure: the Antarctic or Arctic Circles seem promising options for a mission to Enceladus, with optimal launch windows of nearly $10$ years.
    \item A $100$ kg spacecraft can be launched using GigaWatt laser technology. The heuristic cost of a flyby mission to Enceladus is loosely estimated to be $\$320$ million using a cost-optimization model, which is smaller than or comparable to flyby missions with similar payloads involving chemical propulsion. With advancements in laser technology and energy generation/storage, the mission cost could decline further.
    \item The light sail architecture may be coupled to chemical propulsion, thereby enabling orbiter or lander missions to be realized at an overall mission cost potentially comparable to analogous missions based exclusively on chemical propulsion.
    \item If spacecraft miniaturization proceeds as per current forecasts, it might be viable to send a flotilla of hundreds of $\sim 1$ kg SmallSats, to survey the icy moons at a total cost lesser than the New Horizons mission.
\end{itemize}

This paper does not address all facets of a putative laser sail mission to Enceladus and Europa, since that would not be practical. Some of the central aspects that warrant subsequent investigation include: (1) selection and actualization of scientific instruments (beyond the mass spectrometer and camera); (2) guidance, navigation, and control of laser sails (especially in the vicinity of ocean worlds); (3) design and implementation of power and communication subsystems; and (4) a rigorous cost analysis of the mission.

Notwithstanding the additional research needed, our study suggests that astrobiology precursor missions to ocean worlds in our Solar System (notably Enceladus and Europa) are not only theoretically feasible but may be rendered advantageous in the near future due to intrinsic benefits of light sails (potential absence of onboard fuel and short flight times). Hence, we conclude by advocating that forthcoming mission concepts by governmental agencies and private enterprises (e.g., \emph{Breakthrough Starshot}) should accord light sail architectures serious consideration.

\section*{Acknowledgements}
We thank the reviewers for their helpful comments, which helped improve the manuscript.

\bibliographystyle{elsarticle-num}

\bibliography{LightSail}

\end{document}